\documentclass[twocolumn,showpacs,preprintnumbers,amsmath,amssymb]{revtex4}
\usepackage{graphicx}
\usepackage{dcolumn}

\begin{document}


\title{Phonons in random alloys: the itinerant coherent-potential approximation}

\author{Subhradip Ghosh, P. L. Leath and Morrel H. Cohen}
\affiliation{
 Department of Physics and Astronomy,
       Rutgers, the State University of New Jersey, 
136 Frelinghuysen Road, Piscataway,New Jersey 08854-8019, USA} 

\begin{abstract}
We present the itinerant coherent-potential approximation(ICPA), an analytic, translationally
invariant and tractable form of augmented-space-based, multiple-scattering 
theory\cite{klgd} in a single-site approximation for harmonic phonons in
realistic random binary alloys with mass and force-constant disorder.
 We provide expressions for quantities needed for comparison
with experimental structure factors such as partial and average spectral
functions and derive the sum rules associated with them. Numerical results are
presented for Ni$_{55}$Pd$_{45}$ and Ni$_{50}$Pt$_{50}$ alloys which serve as 
test cases, the former for weak force-constant disorder and the latter 
for strong. We present results on dispersion curves and disorder-induced widths.
Direct
comparisons with the single-site coherent potential approximation(CPA)
 and experiment are made which provide
insight into the physics of force-constant changes in random alloys. The CPA accounts well
for the weak force-constant disorder case but fails for strong force-constant disorder
where the ICPA succeeds.

\end{abstract}

\maketitle

\section{Introduction}
\label{sec_Intr}

Much research has been carried out over the past few 
decades on the nature of elementary excitations in disordered alloys.
Many aspects of the lattice-vibrational, magnetic and electronic 
excitations in such systems have been intensively studied both theoretically
and experimentally. Of them, the electronic problem has been covered in
most detail in recent times with the emergence of first-principles
techniques which have made it possible for the theories to attain a much higher
degree of accuracy and reliability. Surprisingly, this is not true
for phonons despite their being not only conceptually
the simplest type of elementary excitation but also the most readily
accessible to detailed experiment. From the early 60's till the early
80's there were many experimental investigations of phonons in
random binary alloys \cite{expt,expt1,expt2,expt3,expt4} by neutron scattering
techniques. More recent experiments have been lacking, probably due to the absence of
a reliable theory. The feature which makes the theory of phonon excitations 
difficult is the inseparability of diagonal and off-diagonal
disorder. The reason for this is that the force-constant sum rule, i.e. the force
constants between a site $i$ and its neighbors $j$ obey the relation 
${\bf \Phi}_{ii}= - \sum_{j\neq i} {\bf \Phi}_{ij}$, must be rigorously satisfied 
even if the system is disordered. In other words, a
single defect at one site in the system perturbs even the diagonal hamiltonian on its neighbors as well,
thereby imposing environmental disorder on the force-constants.
Hence, any theory must include diagonal, off-diagonal, and 
{\it environmental} disorder as well in order to produce reliable results for phonon excitations
in random alloys. 

From the late 60's there were many attempts to provide an 
adequate theory of phonons in random alloys. The first successful, self-consistent 
approximation was the coherent potential approximation(CPA) \cite{cpa}.
The CPA is a single-site, mean-field approximation generally capable of dealing only with diagonal
disorder ({\it mass disorder} in the context of phonons). In the early 70's there were
several studies using the CPA \cite{cpa1,cpa2} which failed to establish it as 
a complete answer to the phonon problem in random alloys. The discrepancies with
experiment confirmed this need for a theory which could include {\it force-constant changes}
in addition to mass disorder. Several extensions of the CPA to include off-diagonal and
environmental disorder were proposed over the next several years \cite{cpa3,cpa4,cpa5,cpa6,cpa7,cpa8} 
but only in certain very special cases, such as the {\it separable}
\cite{cpa3} or the {\it additive} \cite{cpa4,cpa5} limits of off-diagonal and environmental
disorder, were there successes. The more general approximations \cite{cpa6,cpa7,cpa8}
produced Green's functions which either failed to retain the necessary analytic properties, the 
translational invariance of the averaged system, or were not fully self-consistent.
Moreover, all of these extensions failed to capture the effects of multisite or cluster
scatterings which give rise to additional structures in quantities such as the spectral
density functions. Later attempts which met with some success for real alloy systems included
the recursion method \cite{hhk} which can handle large clusters and treats all kinds of disorder
on an equal footing. However, the recursion method is neither self-consistent nor
translationally invariant when used alone. 
Yussouf and Mookerjee \cite{mook} were able to provide a self-consistent generalization of the
CPA to include 2-site scattering using a recursion method in conjunction with the {\it augmented
space formalism}(ASF)\cite{asf}. The ASF has the proper translational invariance, yields
analytic Green's functions, and can handle diagonal and off-diagonal disorder.

An alternative approach was provided by Kaplan, Leath, Gray and Diehl \cite{klgd}(KLGD) which
is also based on the ASF. This approach generalized the {\it travelling cluster approximation}
of Mills and Ratanavararaksha \cite{tca} for diagonal disorder to include the other kinds
of disorder and multisite effects. Using the diagram symmetry rule of Mills and Ratnavararaksha and the
translational symmetry of the augmented-space operators, they presented a self-consistent
multiple-scattering theory which allows one to work with a small number of atoms instead
of treating large clusters as is done in recursion. It provides analytic, translationally-invariant
approximations at all concentrations for diagonal, off-diagonal, and
environmental disorder. It can be applied even to problems of charge transfer,
lattice relaxation, and short-range order in the context of electronic excitations. However,
they illustrated their method only with one-dimensional models and presented it in a
very general and complex mathematical language.

In this paper, we present a simple, straightforward formulation of the KLGD method 
for single-site
scattering of phonons in three-dimensional lattices and provide the first
application of it to phonons in random alloys .
We term it the {\it itinerant coherent-potential approximation} or ICPA ; it maintains
translational invariance, unitarity, and analyticity of physical properties while including
off-diagonal and environmental disorder.
In addition to demonstrating its superiority over the single-site CPA and its previous extensions, 
we provide insight into the physics of force-constant disorder. Our results reveal the complex 
interplay of forces between various atomic species in a random environment, an important
phenomenon which has never been addressed properly.

In Section II we describe the theory, introducing the augmented-space
representation and its use in constructing  the self-consistent scattering theory and
the single-site itinerant coherent-potential approximation. In
Section III we derive expressions for important physical quantities such as densities of states,
spectral functions, inelastic scattering cross-sections and their sum rules 
in terms of the configuration-averaged Green's function of the system. In
Sections IV and V we present our results on Ni$_{55}$Pd$_{45}$ and Ni$_{50}$Pt$_{50}$ alloys
as test cases and compare them with experiment. Concluding remarks are presented in Section VI.
\section{Formalism}
\label{sec_Form}

In this section, we briefly sketch the rationale behind augmented space, introduce
its representations, and define the notation to be used throughout the paper.
We present our discussions here only in the context of phonons. The formulation of the ICPA 
for other kinds of excitations is closely analogous.

\subsection{Augmented space and its representations}

The description of disordered systems conventionally proceeds as follows: the
dynamical behavior of a system is described by a Hamiltonian, whereas
the statistical behavior of the disorder is imposed from outside. 
The Hamiltonian itself
does not describe the full behavior of the random system, but has to be augmented with the
distribution of the set of random potentials which are associated with the various configurations of the
system. The physical properties are then obtained by ensemble averages over
configurations. The CPA and its extensions employ this procedure.

An alternative procedure is that instead of looking at the excitations of the system as moving in
a random array of disordered potentials, the excitations are considered to be moving in periodic
potentials in the presence of a `field' which specifies the disorder. The Hamiltonian, expanded to
include the disorder field, then by itself completely describes the disordered system. 
Since the information on random 
configurations is already incorporated into the Hamiltonian, the configuration averaging is not
a further process as in the mean-field approaches, but simply an evaluation of matrix elements. 
The idea of introducing a `disorder field' to
describe the random fluctuations in the system by extending the Hilbert space to include
the disorder field and by representing the Hamiltonian in this new space constitutes the core of the
augmented-space formalism. The extended Hilbert space which captures the random fluctuations is
called the `augmented space'.

Here, we
work only with a binary alloy A$_{c_{A}}$B$_{c_{B}}$. We assume that each lattice site is
randomly occupied by an A atom or by a B atom. 
We wish to calculate the configuration-averaged values of the experimentally measurable 
physical quantities, for which we need a configuration-averaged Green's function. 
In particular, we shall concentrate here on
the configuration-averaged displacement-displacement (one-phonon) Green's function \cite{rmp}

\begin{equation}
\ll G_{nm}^{\alpha \beta}\left(t \right) \gg = \frac{1}{ih}\ll u_{n}^{\alpha}\left(t \right) ; u_{m}^{\beta}\left(0 \right) \gg ,
\end{equation}
or, after Fourier transformation to the frequency domain
\begin{equation}
\ll {\bf G}\left(\omega^{2} \right) \gg = \ll [{\bf m}\omega^{2} - {\bf \Phi} ]^{-1} \gg .
\end{equation}
In Eqs.(1) and (2) $\ll\enskip \gg$ stands for both configuration and thermodynamic averaging. 
In Eq.(1), $m,n$ specify lattice sites and
$\alpha \beta$ the cartesian directions. $u_{n}^{\alpha}(t)$
is the displacement operator of an atom at the lattice site $n$ in the 
direction $\alpha$ at the time $t$.
In Eq.(2) a bold symbol represents a matrix for which all indices are to be
understood.
The semi-colon $;$ denotes Bose time ordering.
${\bf m}$ is the mass operator, ${\bf \Phi}$ is the force-constant operator, and $\omega$ is
the frequency which contains a vanishingly small negative imaginary part.
The masses are random,
\begin{equation}
m^{\alpha \beta}_{ij} = m_{i}\delta_{\alpha \beta}\delta_{ij},
\end{equation}
with $m_{i}$ randomly taking on the value $m^{\Gamma}$ if species $\Gamma$=A,B
is on site $i$. The force-constants take on the values $\left(\phi^{\alpha\beta}_{ij} \right)^{\Gamma\Delta}$ if species $\Gamma$ is on site $i$ and species $\Delta$ is on site $j$.

It is $\ll {\bf G} \gg$ which carries all the dynamical informations of interest,
and the essential difficulty of the theory of phonons in random systems arises
from taking the configuration average of the inverse of the matrix ${\bf m}\omega^{2}
- {\bf \Phi}$. The augmented-space
technique \cite{asf,km} greatly facilitates this averaging.
The displacements ${\bf u}$, masses ${\bf m}$, force-constants ${\bf \Phi}$, and Greens
function ${\bf G}$ are defined in the dynamical Hilbert space $\Psi$ in which the 
Hamiltonian of the system operates. For a binary alloy, $\Psi$ is augmented by the space
$\Theta$ of all possible atomic configurations of the system. The resulting augmented
space $\Omega$ is
\begin{eqnarray*}
\Omega = \Psi \otimes \Theta ,
\end{eqnarray*}
In $\Omega$ or $\Theta$ operators are represented by symbols with superposed carets.
In the {\it configuration representation} within $\Theta$, the state of site $i$ is
specified by the single-site state $\vert A_{i} \rangle$ if A is on $i$ and by
$\vert B_{i} \rangle$ if B is on $i$. With respect to these states, the occupation operators
$\widehat{\eta}_{i}^{\prime \Gamma}$, $\Gamma$=A,B,
\begin{eqnarray}
\widehat{\bf \eta}_{i}^{\prime A}\vert A_{i} \rangle &=& \vert A_{i} \rangle, \enskip \widehat{\bf \eta}_{i}^{\prime A}\vert B_{i} \rangle = 0 , \nonumber \\
\widehat{\bf \eta}_{i}^{\prime B}\vert B_{i} \rangle &=& \vert B_{i} \rangle, \enskip \widehat{\bf \eta}_{i}^{\prime B}\vert A_{i} \rangle = 0 
\end{eqnarray}
are represented by the matrices
\begin{equation}
\widehat{\eta}_{i}^{\prime A}= \left( \begin{array} {cc}
	1 & 0 \\ 0 & 0
	\end{array} \right) ,
\widehat{\eta}_{i}^{\prime B}= \left( \begin{array} {cc}
	0 & 0 \\ 0 & 1
	\end{array} \right) = \widehat{\bf I}_{i}-\widehat{\eta}_{i}^{\prime A}.
\end{equation}
The configuration of the entire system is specified by the direct product of all
single-site states $\prod_{i} \vert \Gamma_{i} \rangle$, $\Gamma$=A,B. The mass operator for
site $i$ is given by,
\begin{equation}
\widehat{\bf m}_{i}^{\prime}= m^{A} \widehat{\eta}_{i}^{\prime A} + m^{B} \widehat{\eta}_{i}^{\prime B} .
\end{equation}
Similarly the force-constants for sites $i$ and $j$ are given by
\begin{eqnarray}
\widehat{\bf \Phi}_{ij}^{\prime} & = & \phi_{ij}^{AA} \widehat{\eta}_{i}^{\prime A} \widehat{\eta}_{j}^{\prime A}+
\phi_{ij}^{AB} \widehat{\eta}_{i}^{\prime A} \widehat{\eta}_{j}^{\prime B} \nonumber \\
& & \text{\hspace{2.0cm}}+ \phi_{ij}^{BA} \widehat{\eta}_{i}^{\prime B} \widehat{\eta}_{j}^{\prime A} +
\phi_{ij}^{BB} \widehat{\eta}_{i}^{\prime B} \widehat{\eta}_{j}^{\prime B}, 
\end{eqnarray}
with the cartesian indices understood.

Consider now a rotated representation for site $i$ in which the basis vectors for its configuration
space are given by
\begin{eqnarray}
\vert 0_{i} \rangle &=& \sqrt{c_{A}} \vert A_{i} \rangle + \sqrt{c_{B}} \vert B_{i} \rangle , \nonumber \\
\vert 1_{i} \rangle &=& \sqrt{c_{B}} \vert A_{i} \rangle - \sqrt{c_{A}} \vert B_{i} \rangle . 
\end{eqnarray}
Constructing the configuration average of any operator $\widehat{\bf A}$ in $\Theta$ can
be carried out simply by taking the expectation value of $\widehat{\bf A}$ with the state
\begin{equation}
\vert f \rangle = \prod_{i} \vert 0_{i} \rangle ,
\end{equation}
Thus $\vert 0_{i} \rangle$ is the site-average state (or the virtual-crystal state), $\vert 1_{i} \rangle$ describes
a fluctuation away from the average state on site $i$, and
\begin{equation}
\vert f_{i} \rangle = \vert 1_{i} \rangle \prod_{j\neq i} \vert 0_{j} \rangle .
\end{equation}
is the state in which there is a fluctuation or a defect in the average state $\vert f \rangle$
only on site $i$. In this {\it fluctuation representation} the occupation operators 
$\widehat{\eta}_{i}^{\prime A}$ and $\widehat{\eta}_{i}^{\prime B}$ 
are transformed to
\begin{eqnarray}
\widehat{\eta}_{i}^{A} &=& \left( \begin{array} {cc}
	c_{A} & \sqrt{c_{A}c_{B}} \\ \sqrt{c_{A}c_{B}} & c_{B}
	\end{array} \right) \\ \nonumber
\widehat{\eta}_{i}^{B}  &=& \left( \begin{array} {cc}
	c_{B} & -\sqrt{c_{A}c_{B}} \\ -\sqrt{c_{A}c_{B}} & c_{A}
	\end{array} \right) .
\end{eqnarray}
In transforming from the configuration representation to the fluctuation representation,
$\widehat{\bf m}^{\prime}$ goes to $\widehat{\bf m}$ and $\widehat{\bf \Phi}^{\prime}$
to $\widehat{\bf \Phi}$, as given by Eqs.(6) and (7), respectively, with the
$\widehat{\eta}^{\prime \Gamma}$ of Eq.(5) replaced by the $\widehat{\eta}^{\Gamma}$ of Eq.(11).
Thus the dynamical operators $\widehat{\bf m}$ and $\widehat{\bf \Phi}$ are not
diagonal with respect to the number of fluctuations or defects in the fluctuation
representation and can create them, destroy them, or, in the case of 
${\phi}_{ij}$, cause them to travel, or `itinerate'.
We refer to the movement of defects induced by the off-diagonal elements of the
$\widehat{\eta}_{i}^{\Gamma}$ as the {\it itineration of fluctuations} to distinguish it from the
{\it propagation} of phonons. However,
these operators are translationally invariant; the randomness in configuration is thus captured by
translationally invariant operators in the configuration space $\Theta$. The $\widehat{\eta}^{\Gamma}$
operators constitute the disorder field referred to above.

Any operator $\hat{\bf A}$ in this augmented space can be represented in block form,
\begin{equation}
\hat{\bf A}= \left( \begin{array} {cc}
	\bar{\bf A} & {\bf A}^{\prime} \\ {{\bf A}^{\prime}}^{\dagger} & \tilde{\bf A}
	\end{array} \right) ,
\end{equation}
where the bold notation {\bf A} implies a matrix in the site and cartesian indices. The four elements
of the block matrix are given by
\begin{eqnarray}
\bar{\bf A}&=&{\bf P \hat{A}P}, \nonumber \\
{\bf A}^{\prime}&=&{\bf P \hat{A} \left(1-P \right)}, \nonumber \\
{{\bf A}^{\prime}}^{\dagger}&=&{\bf \left(1-P \right) \hat{A} P}, \nonumber \\
\tilde{\bf A}&=&{\bf \left(1-P \right) \hat{A} \left(1-P \right)},
\end{eqnarray} 
where ${\bf P}$, the projection 
operator onto the virtual-crystal state, is given by ${\bf P}= \vert f \rangle \langle f \vert$ .
Thus,we see that $\bar{\bf A}$ is the configuration average of the quantity $\hat{\bf A}$
while ${\bf A}^{\prime}$,${{\bf A}^{\prime}}^{\dagger}$ generate the coupling between the
average and the fluctuation states and $\tilde{\bf A}$ is that part of ${\bf A}$
entirely within the space of fluctuation
states. 

In the present paper we shall make the approximation of 
treating explicitly only single fluctuation
states $\vert f_{l} \rangle$ in the fluctuation space $\Theta - \vert f \rangle \langle f \vert$,
although multiple-fluctuation states are treated implicitly via a self-consistency condition.
States in $\Omega$ can then be specified by $\vert if \rangle$ or $\vert if_{l} \rangle$ where
$i$ is the site index of the dynamical variable in $\Psi$, position or momentum, with the cartesian 
index understood.
For the site indices of the corresponding matrix elements we shall often use the compact notation
\begin{eqnarray}
\langle if \vert \hat{\bf A} \vert jf \rangle & = & \bar{A}_{ij} , \nonumber \\
\langle if_{l} \vert \hat{\bf A} \vert j f_{l^{\prime}} \rangle &=& \tilde{A}^{(l)(l^{\prime})}_{ij}, \nonumber \\
\langle if \vert \hat{\bf A} \vert j f_{l} \rangle & = & A^{{\prime}(l)}_{ij}, \nonumber \\
\langle i f_{l} \vert \hat{\bf A} \vert j f \rangle & = & A^{{\prime}\dagger (l)}_{ij} ,
\end{eqnarray}
where $l$ and $l^{\prime}$ denote the locations of the concentration fluctuation or defect.
The parentheses around $l$ indicate that it is neither a site nor a cartesian direction
index, but indicates instead the position of a {\it fluctuation} in the lattice.

\subsection{Multiple-scattering picture}
A phonon propagating in a random alloy undergoes irreducible multiple scattering \cite{gonis} both repeatedly off a single
fluctuation and successively off fluctuations on the different sites
it encounters in the process. The CPA takes into account the former but not the latter.
To illustrate how the treatment of this process of multiple scattering 
by fluctuations differs between the CPA and our
formalism we employ a cartoon diagram (Fig.1). 
The top panel, a two-dimensional cross-section, illustrates the 
multiple-scattering process included in the CPA. There,
the filled circle is a single `fluctuation site' immersed in an average medium denoted by open
circles. The arrow on the left is the direction of phonon propagation. When the phonon meets the
fluctuation site, it undergoes irreducible multiple scattering at that site. In the
CPA(diagonal disorder), the irreducible scattering by the defect site is 
confined to the defect site. The circle around
the fluctuation site indicates the region of influence of the perturbation. None of the springs
are affected by the presence of this defect since
the force-constants are the same everywhere. One does an averaging over all the possible
occupations of the single site. The phonon diagrams of the self-energy which describe this multiple
scattering process completely are shown in Fig.2(a). There, the filled circles represent the 
fluctuation sites, the dotted lines represent successive scatterings from the fluctuation site, and
the double solid line represents the self-consistent propagator.  
\begin{figure}
\includegraphics[scale=0.6]{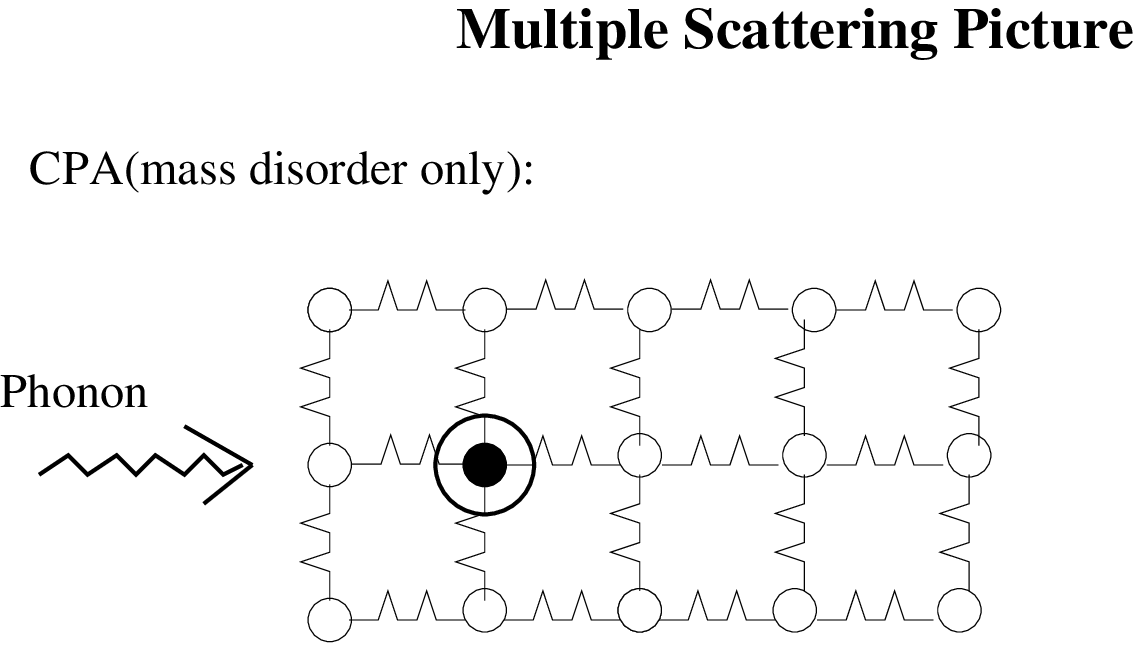}
\vskip 2.0cm
\includegraphics[scale=0.5]{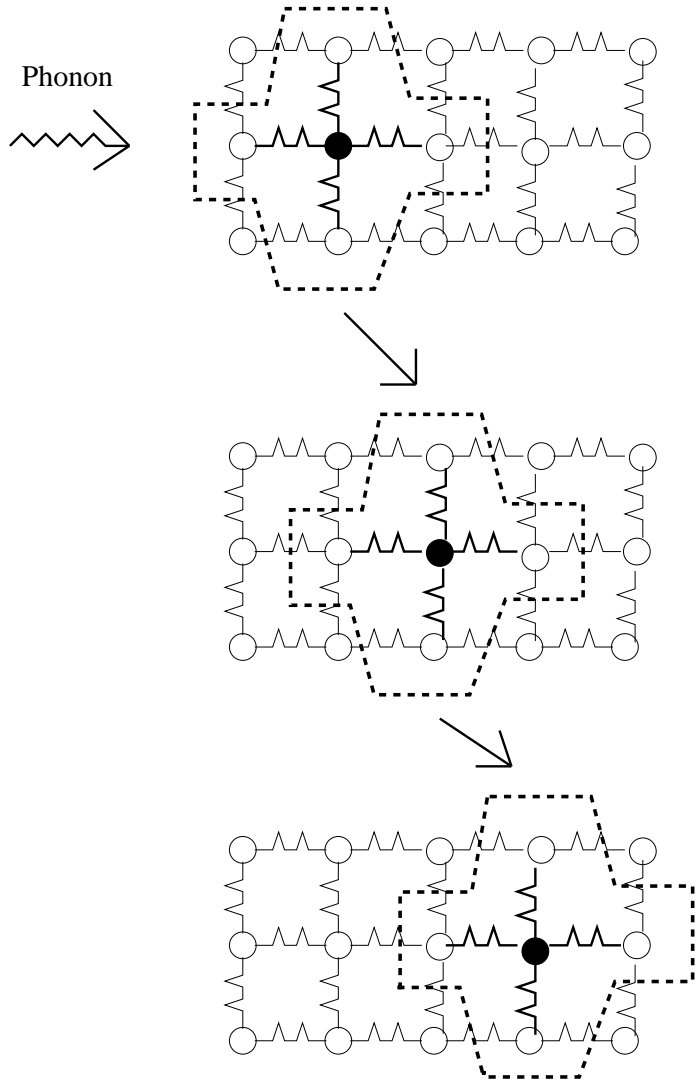}
\caption{Multiple scattering picture in the
CPA(top) and with the ICPA(bottom). The filled circle is the site of the
fluctuation, and the contours around it indicate its area of influence. 
The arrows with the ICPA indicate the {\it itineration} of the
fluctuation to neighboring sites. The details
are given in the text 
.}
\end{figure}
The lower three panels in Fig.1 illustrate scattering sites in the ICPA. The difference 
from the CPA is that the region of influence is not only the site of fluctuation but also its 
neighboring environment around the fluctuation site. The figure shows an example (dotted contour)
where the environment 
includes nearest neighbors only (The calculations could be extended to further neighbors as well). 
When the phonon interacts with
the fluctuation site in the top panel of the three, it scatters also from all of its neighbors since their spring constants 
also undergo changes (denoted by the thick spring lines in contrast to the
thin ones for the average medium). The whole cluster of atoms undergoes fluctuations in 
force-constants as the occupation of the fluctuation site changes. 
One has to keep in mind that the force-constant
between the fluctuation site and its neighbor on the right, say, depends on the occupation of both sites,
as is true for the next neighboring site on its right as well. So, one is led
to include the irreducible scatterings by the fluctuation on all neighboring sites, which then requires inclusion
of scattering by the fluctuations on its neighbors etc., until the irreducible scatterings extend
throughout the entire sample. A simple example of this process is indicated in
the middle and bottom panels. Indeed, Mills and Ratanavararaksha\cite{tca} have shown that once
there are non-diagonal terms in the scattering, the self-energy {\it must} include these migrations
(itinerations) of the scatterer throughout the sample in order to attain unitarity 
and thereby guarantee that the
average Green's function will be properly analytic or Herglotz. The self-consistent scattering and the 
resulting coherent potential about a single defect thus {\it itinerates} from defect to
defect throughout the sample, making it an {\it itinerant coherent potential}. The scattering could have
started from any site in the sample so that the result is also fully translationally invariant, and
the self-energy is $q$-dependent but diagonal in the $q$-space of the Brillouin zone of the 
underlying periodic lattice structure. Fig. 2(b) illustrates a typical self-energy diagram in the
ICPA. The solid and dotted overlapping ellipses denote the multiple 
scattering by a single-site and its neighbors, i.e. by a cluster of atoms, and the subsequent 
itineration of this process. The thin dotted lines and the thick double lines are as in Fig.2(a).
\begin{figure}
\includegraphics[scale=0.4]{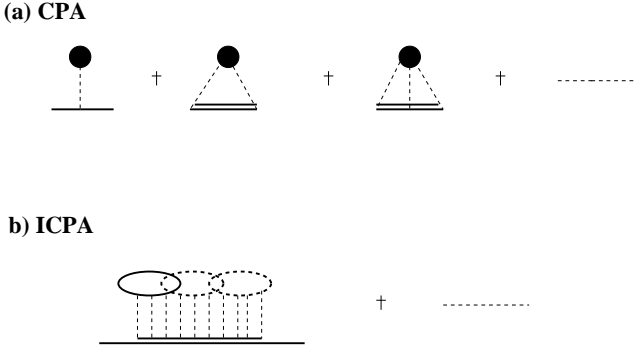}
\caption{The self-energy diagrams in the
CPA(top) and a typical example in the ICPA(bottom). 
The details are given in the text 
.}
\end{figure}

In the multiple scattering framework, we calculate the self-energy ${\bf \Sigma}\left
(\omega^{2} \right)$, defined by
\begin{equation}
\ll {\bf G}\left(\omega^{2} \right) \gg=[{\bf G}_{vca}^{-1}\left(\omega^{2} \right) - {\bf{\Sigma}} \left(\omega^{2} \right) ]^{-1},
\end{equation}
where ${\bf G}_{vca}$ is the unperturbed Green's function, 
\begin{equation}
{\bf G}_{vca} = \left(\overline{\bf m}\omega^{2} -{\overline{\bf{\Phi}}} \right)^{-1},
\end{equation}
and ${\overline{\bf m}}$ and ${\overline{\bf{\Phi}}}$ are the configuration-averaged mass and
force-constant operators respectively.

Our major task is to calculate the self-energy ${\bf{\Sigma}}\left(\omega^{2} \right)$. Let us consider
$\widehat{\bf K}=\left(\widehat{\bf m}\omega^{2} - \widehat{\bf \Phi} \right)=\widehat{\bf G}^{-1}$. Using
the 2$\times$2 block representation of augmented-space operators of Eq.(12) 
we get
\begin{equation}
\widehat{\bf G}= \left( \begin{array} {cc}
	\overline{\bf G} & {\bf G}^{\prime} \\ {{\bf G}^{\prime}}^{\dagger} & \widetilde{\bf G}
	\end{array} \right)
= \left( \begin{array} {cc}
	\overline{\bf K} & {\bf K}^{\prime} \\ {{\bf K}^{\prime}}^{\dagger} & \widetilde{\bf K}
	\end{array} \right)^{-1} .
\end{equation}
Using the relation for the inverse of an operator in 2$\times$2 block form \cite{mat}, namely,
\begin{equation}
{\hat{\bf A}}^{-1}=
\left( \begin{array} {cc}
	\left(\bar{\bf A}-{\bf A}^{\prime}{\widetilde{\bf A}}^{-1}{\bf A}^{\prime \dagger} \right)^{-1} & - \left( \widetilde{\bf A}{{\bf A}^{\prime}}^{-1}\bar{\bf A}- {\bf A}^{\prime \dagger} \right)^{-1} \\
- \left( \bar{\bf A}{{\bf A}^{\prime \dagger}}^{-1} \widetilde{\bf A} -{\bf A}^{\prime} \right)^{-1} & \left( \widetilde{\bf A} -{\bf A}^{\prime \dagger}{\bar{\bf A}}^{-1}{\bf A}^{\prime} \right)^{-1}
	\end{array} \right) ,
\end{equation}
we get
\begin{eqnarray}
\overline{\bf G} & = & [\left(\overline{\bf m}\omega^{2} - \overline{\bf \Phi} \right) - {\bf K}^{\prime} \left(\widetilde{\bf m}\omega^{2} - \widetilde{\bf \Phi} \right)^{-1}{\bf K}^{\prime \dagger} ]^{-1} \nonumber \\
& = & [{\bf G}_{vca}^{-1} -{\bf K}^{\prime} \{{\bf G}_{vca}^{-1} - [\left(\overline{\bf m}\omega^{2} -\overline{\bf \Phi} \right) - \nonumber \\
& & \enskip \enskip \enskip \enskip \enskip \enskip \left(\widetilde{\bf m}\omega^{2} -\widetilde{\bf \Phi} \right) ] \}^{-1} {\bf K}^{\prime \dagger} ]^{-1} \nonumber \\
& = & [{\bf G}_{vca}^{-1} -{\bf K}^{\prime}{\bf F}{\bf K}^{\prime \dagger} ]^{-1} .
\end{eqnarray}
Therefore, the self-energy is given by
\begin{equation}
{\bf{\Sigma}}={\bf K}^{\prime}{\bf F}{\bf K}^{\prime \dagger}, 
\end{equation}
where
\begin{equation}
{\bf F} = \widetilde{\bf K}^{-1}=\{ {\bf G}_{vca}^{-1}\tilde{\bf I}- \widetilde{\bf V} \}^{-1} ,
\end{equation}
and where
\begin{equation}
\widetilde{\bf V}=\left(\overline{m}\tilde{\bf I}-\widetilde{\bf m}\right)\omega^{2}-\left(\bar{\phi}\tilde{\bf I}-\tilde{\bf \Phi} \right).
\end{equation}
The quantity $\widetilde{\bf V}$ denotes all perturbations to the average medium, 
and ${\bf F}$ contains the
itineration of the fluctuation in the average medium. 

Up to this point, the scattering formalism is exact. We now introduce 
the ICPA by restricting the
states within the configuration space $\Theta - \vert f \rangle \langle f \vert$ 
to the single-fluctuation states, the notation
for which is given by Eq.(14). Making the site and cartesian indices explicit,
we obtain for ${\bf \Sigma}$ in Eq.(20), under this restriction,
\begin{equation}
\Sigma_{ij}^{\alpha \beta}=  \sum K^{(m)}_{\alpha i,\delta k}  F^{(m)(n)}_{\delta k,\gamma l} K^{\prime \dagger (n)}_{\gamma l, \beta j} .
\end{equation}
The sumations are over the repeated indices, and the {\it fluctuation itinerator} ${\bf F}$ is given by
a Dyson equation,
\begin{equation}
{\bf F}^{(i)(j)} = {\bf G}_{vca}[ \delta_{(i)(j)} + \sum_{l} \widetilde{\bf V}^{(i)(l)} {\bf F}^{(l)(j)} ] ,
\end{equation}
where only the site index of the fluctuation is shown.
The quantities in (23) are translationally invariant as follows:
\begin{eqnarray}
K^{(m)}_{ik} & = & K^{(0)}_{i-m,k-m} ,\nonumber \\
F^{(m)(n)}_{kl} & = & F^{(0)(n-m)}_{k-m,l-m} .
\end{eqnarray}

The single fluctuation in Eq.(23) can be considered to have been `created' by ${\bf K}^{\prime (n)}$ 
at site $n$, itinerated to site $m$ by ${\bf F}^{(n)(m)}$ and `destroyed' by ${\bf K}^{\prime \dagger (m)}$
at site $m$. The ${\bf K}$, ${\bf K}^{\prime \dagger}$ and ${\bf F}$ matrices have elements which are
non-zero only for site indices within the environment of the appropriate defects i.e. the indices 
$i$ and $k$ ($l$ and $j$) must be within the neighborhood perturbed by the defect at $m$($n$). 
The terms with more than one
fluctuation(defect) present at a time correspond to coherent pair and `defect cluster' scattering and are
neglected in the single-site scattering considered here.
All of these operators act in the augmented space. The Equations (20)-(24)
define an itinerant single-site multiple scattering theory.

\subsection{Self-consistency}

The restriction in Eq.(23) to states of $\Theta- \vert f \rangle \langle f \vert$ containing
only a single fluctuation is a very severe approximation. Multiple-fluctuation states are
of course present in ${\bf F}$ and contribute to ${\bf \Sigma}$. 
In the spirit of the CPA, these are included approximately by introducing self-consistency.
As in the CPA \cite{klgd,rmp} we obtain self-consistency by
replacing ${\bf G}_{vca}$ in ${\bf F}$ in Eq.(24) by a conditional propagator ${\bf G}^{(i)}$,
identical to $\overline{\bf G}= \ll {\bf G} \gg$ except that all irreducible scatterings beginning or
ending on site $i$ are omitted, so that ${\bf F}$ would then be given by
\begin{equation}
{\bf F}^{(i)(j)}={\bf G}^{(i)}[\delta_{(i)(j)}+ \sum_{l}\widetilde{\bf V}^{(i)(l)}{\bf F}^{(l)(j)}] .
\end{equation}
In parallel with Eq.(15), ${\bf G}^{(i)}$ contains a conditional self-energy 
${\bf \Sigma}^{(i)}$
which, is like Eq.(23), except that it includes only those scatterings that neither start nor end on $i$,
\begin{equation}
{\bf G}^{(i)}=[\left({\bf G}_{vca} \right)^{-1} - {\bf \Sigma}^{(i)}]^{-1} , 
\end{equation}
\begin{equation}
{\bf \Sigma}^{(i)}=\sum_{l,m \neq i} {\bf K}^{\prime (l)}{\bf F}^{(l)(m)}{\bf K}^{\prime \dagger (m)} .
\end{equation}
Referring to Fig.2a, the double line in the multiple-scattering graphs is the propagator
${\bf G}^{(i)}$ when the solid dot refers to site $i$.
We obtain the {\it itinerant} CPA by allowing ${\bf K}^{\prime}$,${\bf K}^{\prime \dagger}$,
and $\widetilde{\bf V}$ to include force-constant disorder as well and therefore defect
itineration in Equations (26)-(28).
This closed set of equations defines our single-site, self-consistent,
multiple-scattering theory which, when solved, yields ${\bf F}$. Inserting ${\bf F}$ into
Eq.(20) for ${\bf \Sigma}$ and the result into Eq.(15) then yields ${\bf G}$. It is
already known that Eqs.(15),(20),(26)-(28) have a unique solution which yields a Herglotz
average Green's function \cite{klgd}.
A major difference between this and previous generalizations of the CPA is that for  
scattering from single-site fluctuations 
with off-diagonal and/or environmental disorder, as is considered here, the matrix 
representation of the operator $\widetilde{\bf V}$ has elements which transfer or itinerate
the fluctuation from site to site. This feature causes the self-energy to have
nonzero off-diagonal elements in real-space extending across the sample and thus 
contributes importantly to such
quantities as the two-particle vertex corrections in a way the CPA cannot \cite{chit}.

It now remains to solve these equations, making use of the translational symmetry of
the augmented-space operators. We accomplish this by Fourier transforms on the fluctuation-site 
labels, 
\begin{equation}
A \left(\vec{q} \right)_{mn} = N^{-1} \sum_{l,l^{\prime}} A_{l+m,l^{\prime}+n}^{(l)(l^{\prime})} e^{-i\vec{q}\cdot \vec{R}_{ll^{\prime}}} ,
\end{equation}
and
\begin{equation}
A^{(l)(l^{\prime})}_{l+m,l^{\prime}+n} = N^{-1} \sum_{\vec{q}} A\left(\vec{q}\right)_{mn} e^{i\vec{q}\cdot \vec{R}_{ll^{\prime}}} ,
\end{equation}
where $\vec{R}_{ll^{\prime}}$ is the lattice vector connecting the fluctuation sites $l$ and $l^{\prime}$, 
$m$ and $n$ are neighbors of $l$ and $l^{\prime}$ respectively, and
the $\vec{q}$ sum is over the Brillouin zone.

We can also effect Fourier transforms on the site indices themselves. That of
the self-energy is
\begin{equation}
\Sigma\left(\vec{q} \right) =N^{-1} \sum_{ij} \Sigma_{ij} e^{-\vec{q}\cdot \vec{R}_{ij}} .
\end{equation}
From Eqs.(20), (25) and (29),it follows that
\begin{equation}
\Sigma \left(\vec{q} \right) = \sum_{l,m,n,p} K_{lm}^{(0)} F \left(\vec{q} \right)_{mn} K_{np}^{\prime \dagger (0)} e^{-i \vec{q}\cdot \vec{R}_{lp}} .
\end{equation} 
In this notation, Eq (26) becomes
\begin{equation}
F\left(\vec{q} \right)_{mn} = G_{mn}^{(0)}+ \sum_{r p} G_{m r}^{(0)} \widetilde{V}\left(\vec{q} \right)_{rp} F\left(\vec{q} \right)_{pn} .
\end{equation}
The cartesian indices here are implicit so that each quantity is a 3$\times$3 matrix .

Since the range of interaction in real-space is finite, the perturbations ${\bf K}^{\prime (i)}$ and
$\widetilde{\bf V}\left(\vec{q} \right)$ are finite matrices, 
nonzero only over a finite set of real sites. 
For example, if we consider nearest-neighbor
perturbation only in a single-site approximation, the $\widetilde{\bf V}\left(\vec{q} \right)$ and ${\bf K}^{\prime}$
are 3$\left(Z+1 \right) \times$ 3 $\left(Z+1 \right)$ matrices where $Z$ is the number of nearest
neighbors. This is the minimum matrix size necessary to exhibit all the impurity modes or states about
each fluctuation site.

In full matrix notation, we obtain
\begin{equation}
{\bf F}\left(\vec{q} \right) = [{{\bf G}^{(0)}}^{-1} - \widetilde{\bf V}\left(\vec{q} \right) ] ^{-1} .
\end{equation}
These matrices , for example, for an fcc lattice, are of dimension 39$\times$39.

In order to evaluate ${\bf G}^{(0)}$ we rewrite Eq. (27) as,
\begin{equation}
{\bf G}^{(0)}=[\left({\bf G}_{vca} \right)^{-1} -{\bf \Sigma}^{(0)}]^{-1}=[\ll {\bf G} \gg^{-1} + \widetilde{\bf \Sigma}^{(0)}]^{-1} ,
\end{equation}
where, $\widetilde{\bf \Sigma}^{(0)}=\left({\bf \Sigma} - {\bf \Sigma}^{(0)} \right)$. 
The conditional self-energy $\widetilde{\bf \Sigma}^{(0)}$
contains {\it only} those scatterings which either start or end with a perturbation
caused by a fluctuation at site $0$.
Thus, to evaluate the self-consistent propagator ${\bf G}^{(0)}$,
we need to know $\ll {\bf G} \gg$.
But $\ll {\bf G} \gg$ is obtained from Eq.(15), which becomes
\begin{eqnarray}
\ll  G \left(\vec{q} \right) \gg &=& [ G_{vca}\left(\vec{q} \right)^{-1} -\Sigma\left(\vec{q} \right)]^{-1}, \nonumber \\ 
\ll  G_{ij} \gg & = & N^{-1}\sum_{\vec{q}} \ll  G \left(\vec{q} \right) \gg e^{-i \vec{q} \cdot \vec{R}_{ij}} .
\end{eqnarray}
After reaching self-consistency by the procedure described below, we use these expressions to
calculate densities of states and spectral functions. 

The conditional self-energy $\widetilde{\bf \Sigma}^{(0)}$ can be broken up into two contributions:

(i) Scattering that starts from a defect at site $0$ and ends at
site $j$. 

(ii) Scattering that starts at $j$ but ends at $0$.

This decomposition results in
\begin{eqnarray}
\widetilde{\bf \Sigma}^{(0)} &=& \sum_{j} [{\bf K}^{\prime (0)} {\bf F}^{(0)(j)} {\bf K}^{\prime \dagger (j)} + {\bf K}^{\prime (j)} {\bf F}^{(j)(0)} {\bf K}^{\prime \dagger (0)}] \nonumber \\
& & \enskip \enskip \enskip \enskip - {\bf K}^{\prime (0)} {\bf F}^{(0)(0)} {\bf K}^{\prime \dagger (0)} .
\end{eqnarray}
The last term is subtracted to avoid overcounting when $j$=$0$.

In a block notation similar to that of Eq.(12), we have
\begin{equation}
\widetilde{\bf \Sigma}^{(0)}= \left( \begin{array} {cc}
	{\bf \Sigma}_{1} & {\bf \Sigma}_{3} \\ {\bf \Sigma}^{\dagger}_{3} & 0
	\end{array} \right) ,
\end{equation}
\begin{equation}
{\bf G}^{(0)}= \left( \begin{array} {cc}
	{\bf G}^{(0)}_{1} & {\bf G}^{(0)}_{3} \\ {\bf G}^{\dagger (0)}_{3} & {\bf G}^{(0)}_{2}
	\end{array} \right) ,
\end{equation}
\begin{equation}
\ll{\bf G}\gg= \left( \begin{array} {cc}
	{\bf G}_{1} & {\bf G}_{3} \\ {\bf G}^{\dagger}_{3} & {\bf G}_{2}
	\end{array} \right) .
\end{equation}
where, for a genaral operator $\hat{\bf A}$, ${\bf A}_{1}$ begins and ends with scattering about site $0$
, ${\bf A}_{2}$ neither begins nor ends with scattering about site $0$ 
and ${\bf A}_{3}({\bf A}_{3}^{\dagger})$
begins(ends) with scattering at the site $0$ and ends(begins) with scattering about a 
site different from $0$.
The term ${\bf \Sigma}_{2}$ is $0$ since $\widetilde{\bf \Sigma}^{(0)}$
must begin or end at the site $0$. From Eq. (35), we have
\begin{equation}
{\bf G}^{(0)}= \ll {\bf G} \gg \left({\bf I} + \widetilde{\bf \Sigma}^{(0)} \ll {\bf G} \gg \right)^{-1} ,
\end{equation}
which leads to
\begin{eqnarray}
{\bf G}_{1}^{(0)} & = & 
\overline{\bf X} [{\bf I}+\left({\bf \Sigma}_{1} - {\bf \Sigma}_{3}{\bf G}_{2}{\bf \Sigma}^{\dagger }_{3} \right) \overline{\bf X} + {\bf \Sigma}_{3}{\bf G}^{\dagger}_{3}]^{-1} ,
\end{eqnarray}
where
\begin{equation}
\overline{\bf X}=\left({\bf I}+{\bf G}_{3}{\bf \Sigma}^{\dagger }_{3} \right)^{-1}{\bf G}_{1} .
\end{equation}
after a lengthy algebraic analysis which was previously given in Ref.18.

In order to evaluate these expressions \cite{klgd}, we need to calculate four terms: 
${\bf G}_{1}$,${\bf \Sigma}_{1}$,
${\bf G}_{3}{\bf \Sigma}^{\dagger}_{3}$ and ${\bf \Sigma}_{3}{\bf G}_{2}{\bf \Sigma}^{\dagger
}_{3}$. The first term ${\bf \Sigma}_{1}$ is just a finite sum of finite matrices and can be evaluated 
directly, but the other two terms involve sums which range over all sites in the solid and must be evaluated
by Fourier transforms. This is done in the following way,
\begin{eqnarray*}
\left( G_{3}\Sigma^{\dagger}_{3} \right)_{t,t^{\prime}} & = &  \sum_{m} \sum_{r,n,l} \ll G\left(\omega^{2} \right) \gg_{t,m+r} K^{\prime (m)}_{m+r,m+n} \nonumber \\
& \times & F^{(m)(0)}_{m+n,l} K^{\prime \dagger (0)}_{l,t^{\prime}} - \sum_{r} \ll G\left(\omega^{2} \right) \gg_{t,r} \widetilde{\Sigma}^{(0)}_{r,t^{\prime}} ,
\end{eqnarray*}
which becomes
\begin{eqnarray} 
\left( G_{3}\Sigma^{\dagger}_{3} \right)_{t,t^{\prime}} & = & \frac{1}{N} \sum_{\vec{q}}\sum_{r}  \ll G\left(\vec{q} \right) \gg  e^{i \vec{q} \cdot \vec{R}_{tr}} M\left(\vec{q} \right)_{r,t^{\prime}} \nonumber \\
& & \text{\hspace{2.0cm}}-\left(G_{1}\Sigma_{1} \right)_{t,t^{\prime}},
\end{eqnarray}
and,similarly,
\begin{eqnarray}
\left(\Sigma_{3}G_{2}\Sigma^{\dagger}_{3} \right)_{t,t^{\prime}} &=&   
\frac{1}{N} \sum_{\vec{q},r,r^{\prime}}M\left(\vec{q} \right)_{tr} \ll G\left(\vec{q} \right) \gg e^{i \vec{q} \cdot \vec{R}_{rr^{\prime}}}M\left(\vec{q} \right)_{r^{\prime},t^{\prime}} \nonumber \\
& & -(\Sigma_{1}G_{1}\Sigma_{1}+ \Sigma_{1}G_{3}\Sigma^{\dagger}_{3}+ \Sigma_{3}G^{\dagger}_{3}\Sigma_{1} )_{t,t^{\prime}} ,
\end{eqnarray}
where
\begin{equation}
M\left(\vec{q} \right)_{r,t^{\prime}}= \sum_{nl} K^{\prime (0)}_{rn} F \left(\vec{q} \right)_{nl} K^{\prime \dagger (0)}_{l t^{\prime}} .
\end{equation}
In these equations,
$0$ is the index of the single fluctuation-site in consideration;
$r,r^{\prime},t,t^{\prime},l,n$ are the neighboring sites of $0$; and
$m,m^{\prime}$ are general sites in the sample. So, it is
clear that one needs to work only on matrices of size 3$\left(Z+1 \right) \times$3 $\left(Z+1 \right)$
and use the Fourier transform of operators to handle the itineration of the fluctuation
throughout the entire sample.
An interesting point to note is that the quantities $G_{3}\Sigma_{3}^{\dagger}$ and $\Sigma_{3} G_{2} \Sigma_{3}^{\dagger}$ represent the scattering and itineration of the disturbance including the effect of the 
off-diagonal and environmental
disorder. In case of diagonal-disorder only, they vanish giving ${\bf G}_{1}^{(0)}={\bf G}_{1}\left({\bf I} + {\bf \Sigma}_{1} \right)^{-1}$, which is the CPA self-consistent propagator, and the self-consistent set of 
equations reduces to the CPA equations. 

The inputs to the self-consistency cycle are ${\bf G}^{(0)}_{start}={\bf G}_{vca}$(or some better guess)
, ${\bf K}^{\prime},
{\bf K}^{\prime \dagger}$ and $\widetilde{\bf V}\left(\vec{q} \right)$. 
The procedures for evaluating the latter three quantities are given in the Appendix.
The cycle consists of the following steps :

\begin{enumerate}
\item calculation of ${\bf F} \left(\vec{q} \right)$ using Eqs.(33) and (34).
\item calculation of $\Sigma\left(\vec{q} \right)$ using Eq.(32).
\item calculation of $\ll G \left(\vec{q} \right) \gg$ and $\ll G \left(\omega^{2} \right) \gg$ using
Eq.(36).
\item calculation of ${\bf G}_{1}^{(0)}$ using Eqs.(42),(43),(44) and (45)
\item If the results of steps 1.-4. are acceptably close to those of the previous cycle,
stop. If not, use as input to step 1 and iterate.
\end{enumerate}

The iterations are done till self-consistency is achieved for each $\vec{q}$-point in the Brillouin zone. 
In the process of achieving self-consistency, one calculates $\ll G \gg$ in both real-space and 
in $\vec{q}$-space; each is needed to obtain densities of states and spectral densities respectively. 
In the next section,
we describe how these are used to calculate physical quantities of interest and discuss 
their significance.

\section{Important quantities; sum rules}
In this section we derive results for important physical quantities such as the densities of states
(partial
and total), spectral densities (partial and total) and inelastic scattering cross sections (coherent
and incoherent). All of these are derivable from the real-space and the $\vec{q}$-space configuration-averaged
Green's function and enable us to make direct comparisons with experimental measurements.

\subsection{Densities of states}
The total density of states for a 3-dimensional system is defined as,
\begin{equation}
\nu\left(\omega \right)=\frac{1}{3 \pi N} {\cal I}m \{Tr \ll {\bf mG}(\omega^{2})\gg \},
\end{equation}
where ${\bf m}$ is the mass matrix, and $N$ is the number of sites.
In augmented space we have,
\begin{eqnarray}
\ll mG \gg_{ii} & =& \langle if \vert \widehat{\bf m}\widehat{\bf G} \vert if \rangle, \nonumber \\
& = & \langle if \vert \widehat{\bf m} \vert if \rangle \langle if \vert \widehat{\bf G} \vert if \rangle \nonumber \\
& & + \langle if \vert \widehat{\bf m} \vert if_{i} \rangle \langle if_{i} \vert \widehat{\bf G} \vert if \rangle , \nonumber \\
& = & \overline{m}\overline{G}_{00} + m^{\prime} \langle if_{i} \vert \widehat{\bf G} \vert if \rangle .
\end{eqnarray}
To evaluate the second term, we use the notation of Eq.(12) for the operators $\widehat{\bf G}$ and $\widehat{\bf K}=\widehat{\bf G}^{-1}$. Then, using Eq.(18), we obtain
\begin{equation}
\langle if_{i} \vert \widehat{\bf G} \vert if \rangle={\bf G}^{\prime \dagger}=-\widetilde{\bf K}^{-1} {\bf K}^{\prime \dagger} \overline{\bf G} =-{\bf F K}^{\prime \dagger} \overline{\bf G}.
\end{equation}
We can, therefore, write
\begin{eqnarray*}
\langle if_{i} \vert \widehat{\bf G} \vert if \rangle = - \sum_{l}\sum_{j,n} F_{ij}^{(i)(l)} K_{jn}^{\prime \dagger (l)} \ll G \gg_{ni}. 
\end{eqnarray*}
Fourier transforming over the fluctuation site according to (29) gives
\begin{eqnarray*}
\langle if_{i} \vert \widehat{G} \vert if \rangle & = & -\frac{1}{N} \sum_{l,j,n} \sum_{\vec{q}} F\left(\vec{q} \right)_{0,j-l} e^{i \vec{q} \cdot \vec{R}_{il}} \\
& & \text{\hspace{2.0cm}} \times K^{\prime \dagger}_{j-l,n-l} \ll G \gg_{ni} .
\end{eqnarray*}
The Fourier transform of $ \ll {\bf G} \gg$ on the real-site index now gives,
\begin{eqnarray*}
\langle if_{i} \vert \widehat{G} \vert if \rangle & = & -\frac{1}{N^{2}} \sum_{l,j,n} \sum_{\vec{q}\vec{q^{\prime}}} F\left(\vec{q} \right)_{0,j-l} e^{i \vec{q} \cdot \vec{R}_{il}} \\
& & \text{\hspace{2.0cm}}\times K^{\prime \dagger}_{j-l,n-l} \ll G (\vec{q^{\prime}} )\gg e^{i \vec{q} \cdot \vec{R}_{ni}} .
\end{eqnarray*}
Finally we obtain,
\begin{eqnarray*}
\langle if_{i} \vert \widehat{\bf G} \vert if \rangle & = & -\sum_{mp} \sum_{\vec{q}} F\left(\vec{q} \right)_{0,m}  K^{\prime \dagger}_{mp} e^{i \vec{q} \cdot \vec{R}_{p}} \ll G (\vec{q})\gg ,
\end{eqnarray*}
where, $m=j-l$, $p=n-l$, the neighboring sites perturbed by the fluctuation.
All the terms on the right hand side have been calculated already in the process of achieving self-consistency.
The evaluation of the average density of states is thus straightforward.

The partial density of states for atoms of type $s$ is given by
\begin{equation}
\nu\left(\omega \right)_{s} =  \frac{m_{s}}{3 \pi N} {\cal I}m \{Tr \ll G(\omega^{2})^{ss} \gg_{ii} \} ,
\end{equation}
where
\begin{equation}
\ll G^{ss} \gg_{ii}  = \ll G^{ss} \gg_{00} = \langle 0f \vert \widehat{\bf \eta}_{0}^{s} \widehat{\bf G} \vert 0f \rangle 
\end{equation}
because of translation invariance.
We thus have
\begin{eqnarray}
\ll G^{s} \gg_{0} =  \ll G^{ss} \gg_{00} & = & \langle 0f \vert \widehat{\bf \eta}^{s}_{0} \vert 0f \rangle \langle 0f \vert \widehat{\bf G} \vert 0f \rangle \nonumber \\
& + & \langle if \vert \widehat{\bf \eta}^{s}_{0} \vert 0 f_{0} \rangle \langle 0 f_{0} \vert \widehat{\bf G} \vert 0f \rangle ,
\end{eqnarray}
and, from Eq. (11), it follows that
\begin{eqnarray}
\nu\left(\omega \right)_{A} & = & -\frac{m_{A}}{3 \pi } {\cal I}m [c_{A} \{\langle 0f \vert \widehat{\bf G} \vert 0f \rangle \} \nonumber \\
& & \enskip + \sqrt{c_{A}c_{B}} \{\langle 0f_{0} \vert \widehat{\bf G} \vert 0f \rangle \} ] , \nonumber \\
\nu\left(\omega \right)_{B} & = & -\frac{m_{B}}{3 \pi } {\cal I}m [c_{B} \{\langle 0f \vert \widehat{\bf G} \vert 0f \rangle \} \nonumber \\
& & \enskip - \sqrt{c_{A}c_{B}} \{\langle 0f_{0} \vert \widehat{\bf G} \vert 0f \rangle \}] . \nonumber \\
\end{eqnarray}
The elements of $\widehat{\bf G}$ in Eq.(53) were already evaluated while 
calculating the average density
of states above.

The partial Green's functions $\ll G^{s} \gg_{0}$ are used in calculating the incoherent scattering structure
factor which is directly measured in the experiments,
\begin{equation}
\ll S_{incoh}\left(\vec{Q},\omega \right) \gg =  \sum_{s} b^{2}_{s} \vec{Q}\cdot {\cal I}m \ll G^{s}\left(\omega \right) \gg_{0} \cdot \vec{Q} ,
\end{equation}
where $b_{s}$ is the incoherent scattering length for atoms of type $s$, and $Q$ is the phonon wavenumber.

\subsection{Spectral densities}
The average spectral function is defined as,
\begin{equation}
\ll {\cal A}_{\lambda}\left(\vec{q},\omega^{2} \right) \gg =\frac{1}{\pi}{\cal I}m \ll G_{\lambda}\left(\vec{q},\omega^{2} \right) \gg ,
\end{equation}
where $\lambda$ is a normal-mode branch index.
More interesting quantities to calculate are the conditional or partial Green's functions 
$\ll G^{ss^{\prime}}\left(\vec{q},\omega^{2} \right) \gg$ in $\vec{q}$-space 
because these enable one to calculate the coherent-scattering structure factors which are measured 
directly in the neutron-scattering experiments and are given by,
\begin{equation}
\ll S_{\lambda}\left(\vec{q},\omega \right) \gg_{coh} = \sum_{ss^{\prime}} d_{s}d_{s^{\prime}}\frac{1}{\pi}{\cal I}m \ll G^{ss^{\prime}}_{\lambda}\left(\vec{q},\omega^{2} \right) \gg ,
\end{equation}
where, $d_{s}$ is the coherent scattering length for the species $s$.

The conditional Green's functions are defined as
\begin{eqnarray*}
\ll G^{ss^{\prime}}\left(\vec{q},\omega^{2} \right) \gg = \frac{1}{N} \sum_{ij} \ll G^{ss^{\prime}}\left(\omega^{2} \right) \gg_{ij} e^{-\vec{q} \cdot \vec{R}_{ij}} ,
\end{eqnarray*}
\begin{eqnarray}
\ll G^{ss^{\prime}}\left(\omega^{2} \right) \gg_{ij} = \ll \widehat{\bf \eta}^{s}_{i} \widehat{\bf G}\left(\omega^{2} \right) \widehat{\bf \eta}^{s^{\prime}}_{j} \gg = \langle if \vert \widehat{\bf \eta}^{s}_{i} \widehat{\bf G} \widehat{\bf \eta}^{s^{\prime}}_{j} \vert jf \rangle \nonumber \\
= \langle if \vert \widehat{\bf \eta}^{s}_{i}\vert  if \rangle \langle if \vert \widehat{\bf G}\vert jf \rangle \langle jf \vert \widehat{\bf \eta}^{s^{\prime}}_{j} \vert jf \rangle \nonumber \\
+ \langle if \vert \widehat{\bf \eta}^{s}_{i}\vert  if_{i} \rangle \langle if_{i} \vert \widehat{\bf G}\vert jf \rangle \langle jf \vert \widehat{\bf \eta}^{s^{\prime}}_{j} \vert jf \rangle \nonumber \\
+ \langle if \vert \widehat{\bf \eta}^{s}_{i}\vert  if \rangle \langle if \vert \widehat{\bf G}\vert jf_{j} \rangle \langle jf_{j} \vert \widehat{\bf \eta}^{s^{\prime}}_{j} \vert jf \rangle \nonumber \\
+ \langle if \vert \widehat{\bf \eta}^{s}_{i}\vert  if_{i} \rangle \langle if_{i} \vert \widehat{\bf G}\vert jf_{j} \rangle \langle jf_{j} \vert \widehat{\bf \eta}^{s^{\prime}}_{j} \vert jf \rangle .
\end{eqnarray}
In Eq.(57) the index $\lambda$ is to be understood.
These four terms include all the possible scattering processes when two different sites are occupied
by two species. The four different terms involve calculations of the Green's function under
various circumstances of coupling between the average and the fluctuation states weighted by the
appropriate concentrations .

We obtain from (57)
\begin{eqnarray}
\ll G^{ss^{\prime}} \gg_{ij} & = & c_{s}c_{s^{\prime}} \langle if \vert \widehat{\bf G} \vert jf \rangle + [c_{s^{\prime}}\sqrt{c_{s}(1-c_{s})}\nonumber \\
& \times & \left(-1 \right)^{(1-n^{s})} \langle if_{i} \vert \widehat{\bf G} \vert jf \rangle] + [c_{s} \sqrt{c_{s^{\prime}}(1-c_{s^{\prime}})} \nonumber \\
& \times & \left(-1 \right)^{(1-n^{s^{\prime}})} \langle if \vert \widehat{\bf G} \vert j f_{j} \rangle] + c_{s}c_{s^{\prime}}\nonumber \\
& & \text{\hspace{1.0cm}} \times \left(-1 \right)^{(n^{s}+n^{s^{\prime}})} \langle if_{i} \vert \widehat{\bf G} \vert jf_{j} \rangle .
\end{eqnarray}
The integer $n^{s}$ is equal to 1 if $s$=A and is equal to 0 if $s$=B.

These terms can be easily calculated using Fourier transforms as has been previously demonstrated 
for the density of states. The final forms of the conditional Green's functions
in $\vec{q}$-space are
\begin{eqnarray}
\ll G^{AA}\left(\vec{q},\omega^{2} \right)\gg & = & c_{A}^{2}\ll G\left(\vec{q},\omega^{2} \right) \gg + c_{A}\sqrt{c_{A}c_{B}}(T_{1}+ \nonumber \\
& & \enskip \enskip \enskip T_{2})+c_{A}c_{B} T_{3} , \nonumber \\
\ll G^{BB}\left(\vec{q},\omega^{2} \right)\gg & = & c_{B}^{2}\ll G\left(\vec{q},\omega^{2} \right) \gg - c_{B}\sqrt{c_{A}c_{B}}(T_{1}+ \nonumber \\
& & \enskip \enskip \enskip T_{2})+ c_{A}c_{B} T_{3} , \nonumber \\
\ll G^{AB}\left(\vec{q},\omega^{2} \right)\gg & = & c_{A}c_{B}\ll G\left(\vec{q},\omega^{2} \right) \gg + \sqrt{c_{A}c_{B}}(c_{B}T_{1} \nonumber \\
& & \enskip \enskip \enskip -c_{A}T_{2} )- c_{A}c_{B} T_{3} , \nonumber \\
\ll G^{BA}\left(\vec{q},\omega^{2} \right)\gg & = & c_{A}c_{B}\ll G\left(\vec{q},\omega^{2} \right) \gg - \sqrt{c_{A}c_{B}}(c_{A}T_{1} \nonumber \\
& & \enskip \enskip \enskip -c_{B}T_{2} )- c_{A}c_{B} T_{3} ,
\end{eqnarray}
where
\begin{eqnarray}
T_{1} & = & \sum_{nm} F_{0n}\left(\vec{q} \right) K^{\prime \dagger}_{nm} e^{i \vec{q} \cdot \vec{R}_{m}} \ll G \left(\vec{q},\omega^{2} \right) \gg ,\nonumber \\
T_{2} & = & \sum_{nm} \ll G\left(\vec{q},\omega^{2} \right) \gg e^{-i \vec{q} \cdot \vec{R}_{n}} K^{\prime}_{nm} F_{m0} \left(\vec{q} \right) ,\nonumber \\
T_{3} & = & F_{00}\left(\vec{q} \right) + \sum_{nm}\sum_{lp} F_{0n} \left(\vec{q} \right) K^{\prime \dagger}_{nm} e^{i \vec{q} \cdot \vec{R}_{m}} \nonumber \\
& & \text{\hspace{0.5cm}} \times \ll G \left(\vec{q},\omega^{2} \right) \gg e^{-i \vec{q} \cdot \vec{R}_{l}} K^{\prime}_{lp} F_{p0}\left(\vec{q} \right) ,
\end{eqnarray}
and where $n,m,l$ and $p$ are the neighboring sites of the fluctuation site $0$ influenced by the 
perturbation. For the lattices with each site having inversion symmetry, $T_{1}$=$T_{2}$ holds
because $T_{1}$ and $T_{2}$ are the contributions from two processes which are conjugate
to one another. In that case, $\ll G^{AB} \left(\vec{q},\omega^{2} \right) \gg$=
$\ll G^{BA} \left(\vec{q},\omega^{2} \right) \gg$ when $c_{A}$=$c_{B}$. 

The sum rules for the conditional Green's functions are derived the following way:
Integrate $\omega\enskip \widehat{\bf G}\left(\omega^{2} \right)$, Eq.(2), along the
real axis, closing the contour above at infinity, obtaining
\begin{eqnarray*}
\oint d\omega \enskip \omega \enskip \widehat{\bf G}\left(\omega^{2} \right)  & = & \widehat{\bf m}^{-1} \pi i ,
\end{eqnarray*}
or 
\begin{eqnarray}
\int_{0}^{\infty} d\omega \enskip 2 \omega \enskip {\cal I}m \widehat{\bf G}\left(\omega^{2} \right)  & = & \widehat{\bf m}^{-1} \pi .
\end{eqnarray}
Similarly, using Eq.(57), taking the Fourier transform of Eq.(58), carrying out the contour
integral and inserting (61) yields the sum rule for the partial spectral functions
\begin{equation}
\int_{0}^{\infty} d\omega \enskip 2 \omega \enskip {\cal I}m \enskip \ll G^{ss^{\prime}}\left(\vec{q},\omega^{2} \right) \gg= \pi \frac{c_{s}}{m_{s}} \delta_{ss^{\prime}} .
\end{equation}
For the total Green's function we obtain
\begin{equation}
\int_{0}^{\infty} d\omega \enskip 2 \omega \enskip {\cal I}m \enskip \ll G\left(\vec{q},\omega^{2} \right) \gg= \pi \left(\frac{c_{A}}{m_{A}}+ \frac{c_{B}}{m_{B}} \right) .
\end{equation}

The experimental dispersion curves are obtained from the wave-vector dependence of the peak
frequencies of the structure factors as measured, after a deconvolution of the experimental resolution
function. The question is whether the dispersion curves so obtained, which incorporate the
effect of the coherent scattering lengths, differ significantly from those obtained from the
peak frequencies of the Green's function itself which gives, in principle, a proper description
of the dynamics but does not contain the scattering length weighting. 
To answer that question one needs to recognize that the peak positions in ${\cal I}m {\bf G}$ are
very closely related to the zeroes of ${\cal R}e {\bf G}^{-1}$ at a given wave vector. 
If we diagonalize the Hermitian ${\cal R}e {\bf G}^{-1}$
both with respect to mode and species index, each of the two species components of ${\cal R}e {\bf G}^{-1}$ will have a zero.
Correspondingly, each of the two components of ${\cal I}m {\bf G}$ will have a peak, if ${\cal I}m {\bf \Sigma}$
does not wipe it out. So, in the species representation, the different matrix elements
${\cal I}m G^{ss^{\prime}}$ will thus all have these peaks at nearly the same frequencies. 
Thus, the weighting
of the $G^{ss^{\prime}}$ by the scattering lengths will not shift the peak positions significantly 
even when the
scattering lengths differ appreciably. The intensities and the lineshapes of $S_{coh}$ and ${\cal I}m {\bf G}$ may
differ significantly, but the peak positions will generally differ little. 
In summary, the structure factor fairly accurately reflects the phonon dynamics contained
in ${\bf G}$ with regard to the dispersion curves, an important fact illustrated below in the next two sections where we
present our calculations on Ni$_{55}$Pd$_{45}$ and Ni$_{50}$Pt$_{50}$ alloys.

\section{Application to Ni$_{55}$Pd$_{45}$; weak force-constant disorder}
In the next two sections we explore the relative importance of force-constant disorder 
and mass disorder for the vibrational properties of random alloys
in two specific alloys Ni$_{55}$Pd$_{45}$ and Ni$_{50}$Pt$_{50}$.
In the former alloy, the mass disorder is much larger than the force-constant disorder.
The mass ratio $m_{Pd}/m_{Ni}$ is 1.812, whereas the Pd force-constants are only about
15$\%$ larger than those of Ni \cite{dbm,bvk}. In the latter alloy, both the mass disorder and
the force-constant disorder are large, providing an interesting contrast between the two
materials. For both the cases we have done our calculations on 200 $\omega$-points and
have used a small imaginary frequency part of -0.01 in the Green's function. For the
Brillouin zone integration 356 ${\vec{q}}$-points in the irreducible 1/48-th of the zone produced well
converged results. The simplest linear-mixing scheme was used to accelerate
the convergence. For both the cases the number of iterations ranged from 3 to 13 depending
on the frequency $\omega$. 

For Ni$_{55}$Pd$_{45}$, we compare the results of virtual crystal(VCA), CPA, and ICPA computations, using
the ICPA force-constants to construct the averages used in the VCA and CPA and compare
the results with experiment. We make a distinction between that use of the VCA and of `mean
crystal' models in which the average mass is employed and a set of `mean-crystal' force-constants
are fitted as parameters to the experimental data.

Kamitakahara and Brockhouse
\cite{expt2} investigated Ni$_{55}$Pd$_{45}$ by inelastic neutron scattering and 
reported a strange
observation. A theoretical calulation based on a mean crystal model having the average mass and 
fitted force-constants between
those of Ni and Pd agreed closely with the experimental dispersion curves. This was quite a 
puzzle because it suggested that the large mass disorder had little effect. There were
theoretical studies on this system using recursion\cite{mook1} and the
average t-matrix approximation \cite{kt}, but no theoretical results for the frequencies were available. 
In an attempt to solve this puzzle, we have carried out calculations with the CPA, the 
ICPA, and the VCA, as well as with the mean-crystal
model used in Ref. 3 .

\begin{figure}
\includegraphics[scale=0.5]{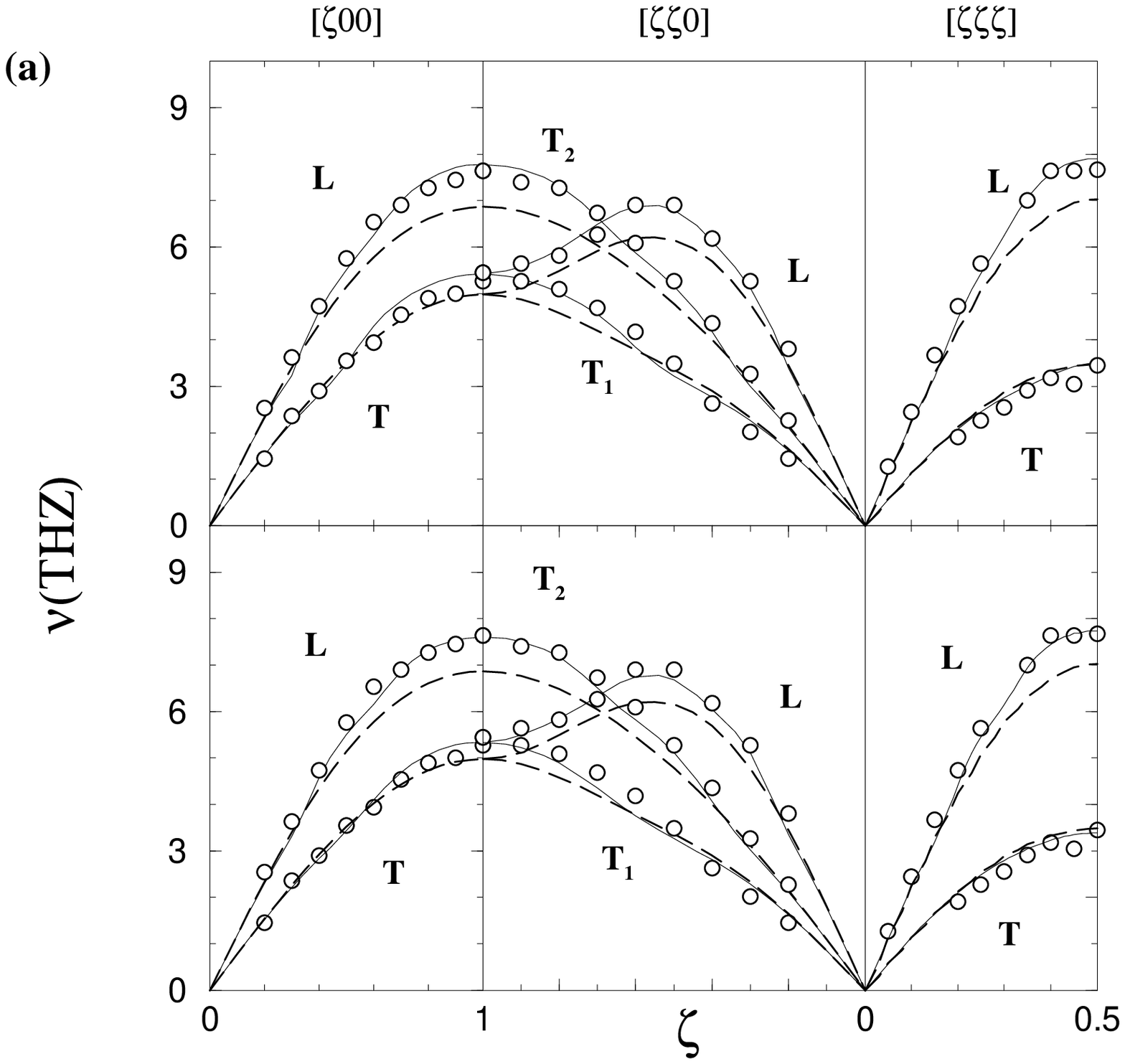}
\includegraphics[scale=0.5]{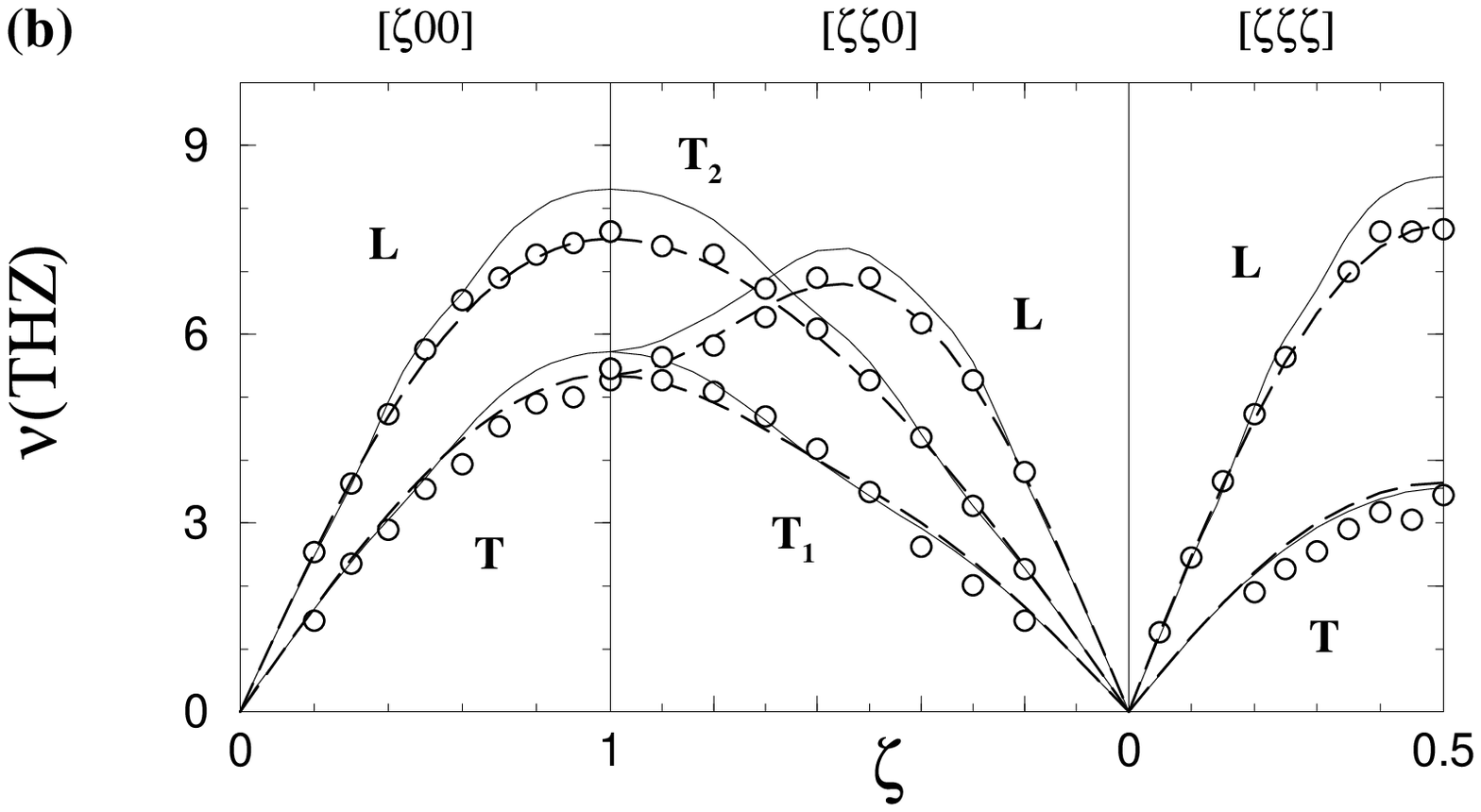}
\caption{(a)(Top panel)Dispersion curves (frequency $\nu$ vs. reduced wave-vector $\zeta$) for
Ni$_{55}$Pd$_{45}$ calculated
in the ICPA(solid line) and in the VCA(dashed line). The circles are the experimental data
\cite{expt2}.
(Bottom panel) Dispersion curves for Ni$_{55}$Pd$_{45}$ calculated in the CPA(solid line) and in the VCA(dashed line). 
The force-constants used are given in the text. The
circles are the experimental data \cite{expt2}.
(b) Dispersion curves for Ni$_{55}$Pd$_{45}$ calculated in the CPA(solid line) and in the mean-crystal model(dashed
line) using the force-constants of Ref. 3. The circles are the experimental data \cite{expt2}.}
\end{figure}
For the ICPA calculation, we assumed that the explicit scattering caused by the force-constant disorder 
was confined to the
nearest-neighbors. This assumption is justified because the nearest-neighbor force-constants are 
an order
of magnitude larger than those of the further neighbors so that the nearest-neighbors feel the effect of
disorder most strongly. For the virtual-crystal or the average medium into which the scattering was embedded,
we kept terms in the Hamiltonian up to the fourth
neighbor, which turned out to be sufficient. The problem with force-constant disorder scattering calculations is the
general absence of prior information about species-dependent force-constants. 
We note that Pd is the larger atom here. In the alloy, the Ni-Pd separation is larger than the
Ni-Ni separation. As a result,
the Ni-Pd force-constants should be less than the Ni-Ni ones. Using this intuitive
argument and for simplicity in this illustration, we kept the $\phi_{Ni-Ni}^{\alpha \beta}$ and 
$\phi_{Pd-Pd}^{\alpha \beta}$ the same as those of the pure
materials \cite{dbm} and reduced the $\phi_{Ni-Pd}^{\alpha \beta}$ below the $\phi_{Ni-Ni}^{\alpha\beta}$
by an $\alpha \beta$ independent factor. The dispersion curves were obtained from the ICPA calculations using
$\phi_{Ni-Pd}^{\alpha\beta}$=0.7$\phi_{Ni-Ni}^{\alpha\beta}$ (solid lines) for the nearest neighbors and using the
force-constants of Ref.3 for the higher neighbors. They are
compared in the top panel of Fig.3(a) with the experimental results \cite{expt2}
(open circles) and the VCA for the same force-constants(dashed lines).
These ICPA  dispersion curves were constructed by numerically determining the peaks in the
coherent scattering structure factor $\ll S_{\lambda}\left(\vec{q},\omega \right) \gg_{coh}$
given by Eq.(56), which was calculated using the partial spectral functions of Eqs.(59) and (60) and
weighting them with the coherent scattering lengths for Ni and Pd. We could thus make
a direct comparison with the experimental results because the neutron data observed in the
experiments inherently incorporates the effect of the scattering lengths of the species.
\begin{figure}
\includegraphics[scale=0.5]{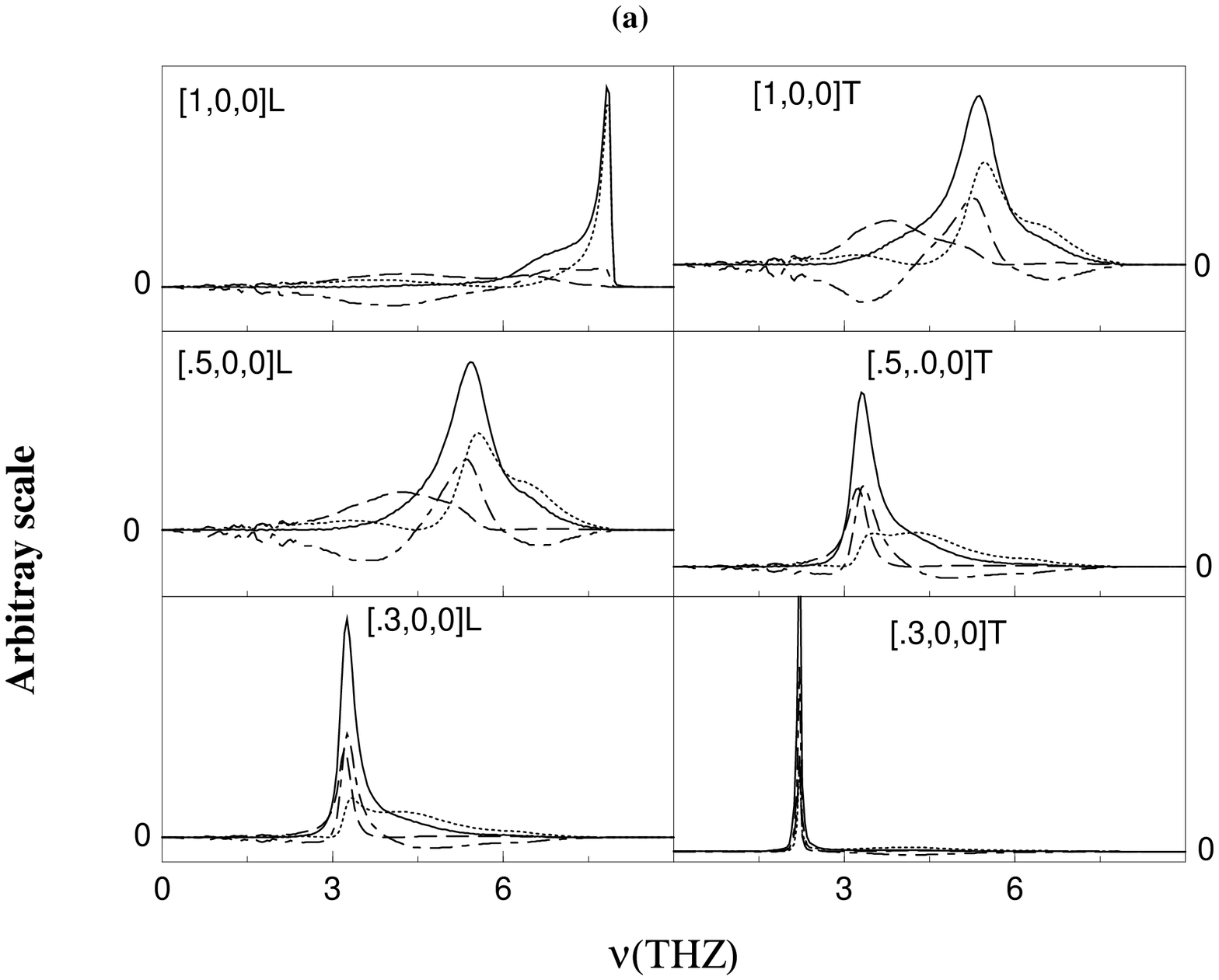}
\includegraphics[scale=0.5]{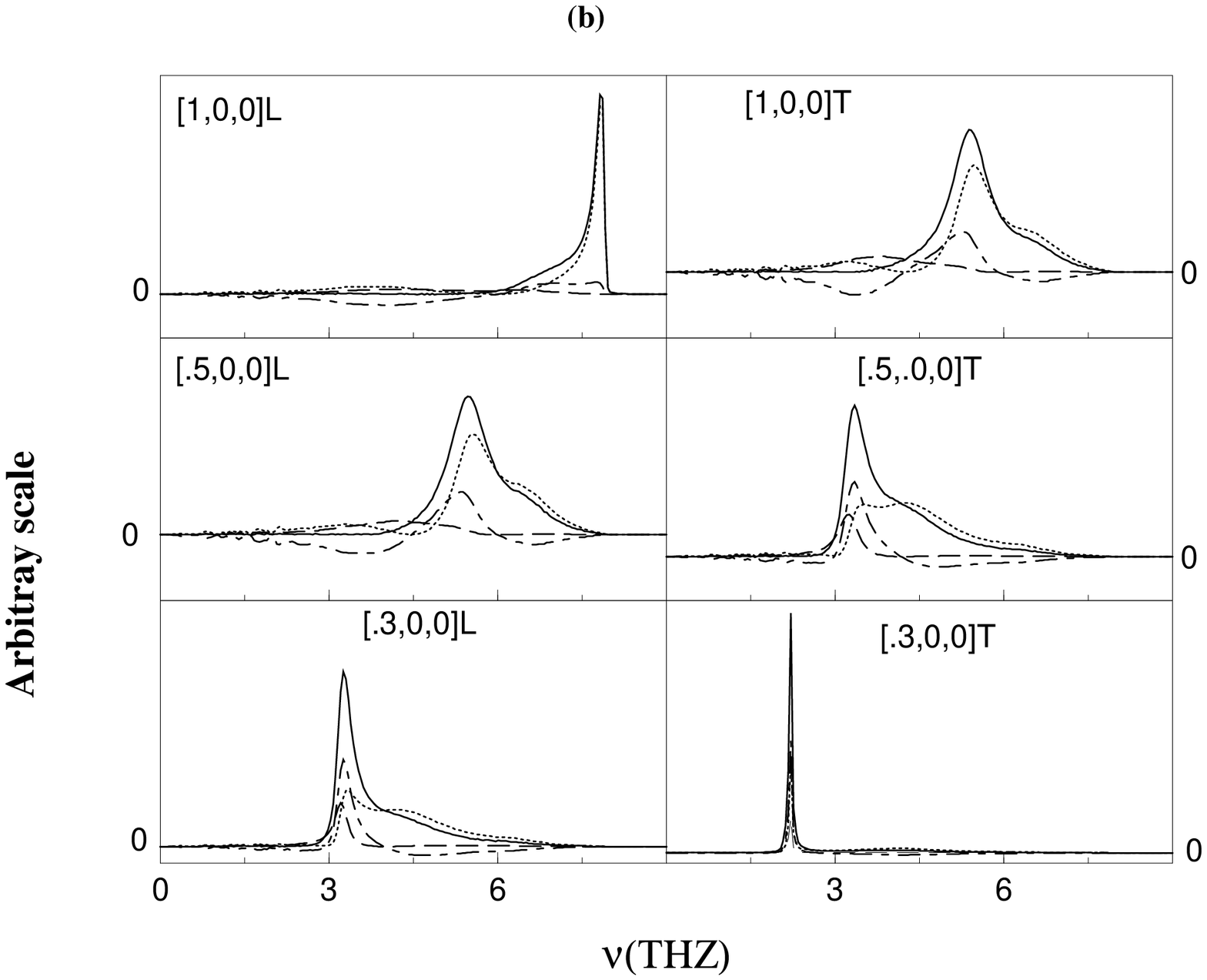}
\caption{(a)Partial and total spectral functions calculated in the ICPA 
for various $\zeta$ values in the [$\zeta$,0,0] direction in Ni$_{55}$Pd$_{45}$. (b) Partial and
total structure factors calculated in the ICPA for various  $\zeta$ values in the [$\zeta$,0,0] direction 
in Ni$_{55}$Pd$_{45}$. The solid lines are the total contribution, the dotted lines are the 
Ni-Ni spectra, the long-dashed lines are the Pd-Pd spectra and the dot-dashed lines are the Ni-Pd
contributions. The details are given
in the text.
The left(right) column is for longitudinal(transverse) modes.}
\end{figure}
Excellent agreement of the ICPA with
experiment was obtained for all three symmetry directions and for each branch
by varying only one parameter in the force-constant
matrix. This suggests that the force-constant disorder is weak and the system is dominated by
the mass-disorder, as one would expect from the numerical values of the parameters. 
It is confirmed by the results of the CPA calculations shown in the 
bottom panel of the Fig. 3(a), using the same force-constants. 
As in the top panel, the solid lines are the CPA results, the circles are the experimental points,
and the dashed lines are the VCA results. The agreement with the experiment 
again suggests the dominance of
the mass-disorder, but there are more interesting points to note. In the long-wavelength (low $\vec{q}$)
regime, the VCA, the CPA, and the ICPA curves are indistinguishable because the self-averaging
of both mass and force-constants over a single wavelength reduces both the CPA and the ICPA to the VCA. 
But, as we move to high wave-vectors , the VCA deviates to
frequencies below the experimentally observed ones. This fact is due to the use of an average mass in the
Hamiltonian. In the high-wave-vector region, the lighter atoms, i.e. Ni in this case, dominate
and push the frequencies up. That is why the CPA
and the ICPA agree very well across the Brillouin zone while the VCA fails for the high wave-vectors. The
reason that Kamitakahara and Brockhouse got a very good fit to the experimental points in Ref. 3 
by using their
mean-crystal model is that they obtained parametrized force-constants which were higher than 
those calculated in the VCA. Though
they had used the average mass in their calculations the higher values of the force-constants
(They used $\overline{\bf \Phi}=c_{A}{\bf \Phi}_{AA}+c_{B}{\bf \Phi}_{BB}$ rather than
$\overline{\bf \Phi}=c_{AA}^{2}{\bf \Phi}_{AA}+c_{BB}^{2}{\bf \Phi}_{BB}+2c_{A}c_{B}{\bf \Phi}_{AB}$) 
compensated for their omission of the effect of the mass fluctuations.
This is illustrated in Fig.3(b). There, the dashed lines are their
mean-crystal model calculations, the circles are the experimental points, and the solid lines 
represent a CPA
calculation with the force-constants used in Ref. 3. Here we see that the CPA yields
frequencies that are too high in the large wave-vector region.
The CPA captures the effect of mass-fluctuation and the domination of the Ni atoms 
for higher wave-vectors, but
the higher values of the assumed mean force-constants pushes the frequencies up further, thereby worsening the agreement with the
experiment. But, there was yet another puzzle to solve. Another striking feature is that in spite of
incorporating the scattering lengths in our calculations in determining the frequencies there was little change in the
ICPA results with respect to the CPA results even though the coherent scattering lengths of Ni and Pd differ
significantly( The coherent scattering length for Ni is 1.03 while that of Pd is 0.6). This lack of change can be
understood from a comparison between the partial and total spectral functions (Fig. 4(a)) and the partial and the
total coherent structure factors (Fig 4(b)). In these figures, we have shown examples of ICPA spectral
functions and structure factors along the [$\zeta$,0,0] direction, $\zeta=\frac{\vert \vec{q} \vert}{\vert \vec{q}_{max} \vert}$,
for a low $\zeta$ , one in the middle of the Brillouin
zone and one at the edge of the Brilluoin zone. In Fig. 5(a), the solid lines are the total spectral function while
the dotted lines are the Ni-Ni spectra, the long-dashed lines are the Pd-Pd spectra and the dot-dashed
lines are the Ni-Pd contribution. In each case, the peaks corresponding to the dominating species and that 
in the Ni-Pd curves occur at the same general positions. For example, in the [.3,0,0]-L curves, the 
peak in the spectral function is mostly that of Pd atoms while for the [.5,0,0]-L and [1,0,0]-L curves,
the contributions are from Ni atoms, the Pd-Pd contribution here is much less and that
too is almost completely neutralised by the Ni-Pd contribution in the low frequency region. The occurence of the
peaks of the Ni-Pd spectral functions and that of the Ni-Ni or Pd-Pd spectral functions almost at the same
position across the Brillouin zone suggests that the inclusion of scattering lengths would primarily alter the relative
weights of various contributions and thereby altering only the line shapes, while the dispersion curves wouldn't
undergo  significant change. This is demonstrated in Fig. 4(b). There, the solid lines are the coherent
structure factors $\ll S_{\lambda}\left(\vec{q}, \omega \right) \gg$, the dotted lines are the Ni-Ni spectral
functions weighted using the scattering length of Ni, the long-dashed lines are the Pd-Pd spectral functions
weighted using the scattering length of Pd and the dot-dashed lines are the Ni-Pd spectral functions weighted
by the scattering lengths of Ni and Pd according to Eq.(56). One can see that the weighting 
affects primarily the peak heights.
These explicit numerical results confirm the qualitative argument given at the end of section III.
\begin{figure}
\includegraphics[scale=0.5]{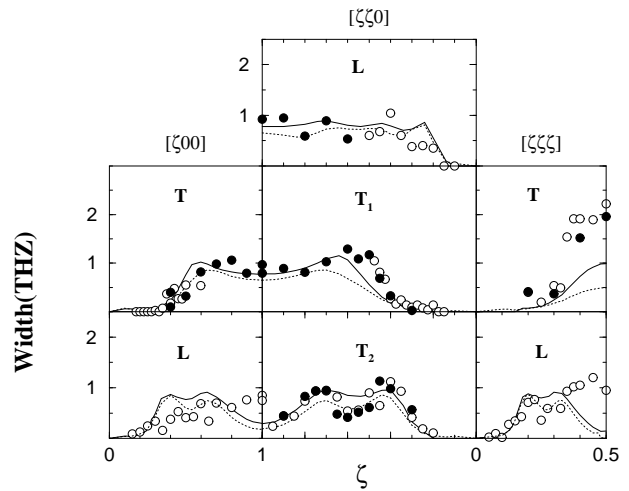}
\caption{Disorder-induced FWHM's in Ni$_{55}$Pd$_{45}$ calculated in the ICPA(solid line) and in the CPA(dotted line)
using the force-constants of Fig. 3(a). The 
circles are the widths extracted from the experimental results \cite{expt2}. 
The filled circles had better experimental resolution.} 
\end{figure}

The disorder-induced widths are important because the effect
of disorder is often manifested in them more directly than in the frequencies. Kamitakahara and Brockhouse
extracted full widths at half maxima(FWHM) from their neutron groups 
by assuming that the observed line shape could be adequately approximated by the convolution of a Gaussian
resolution function(representing the experimental resolution) with a Lorentzian natural line shape. 
Thus, for a comparison of our results
with theirs,
we have fitted our structure factors  to a Lorentzian to extract the widths. The results are shown in Fig.5. The
solid lines are the widths (FWHM) obtained in the ICPA, and the dotted lines are 
those obtained in the CPA. The circles are the experimental points, 
the filled circles having better resolution \cite{expt2} were those in which Kamitakahara and Brockhouse had
more confidence.
Generally, there is
little difference between the widths obtained in the CPA and the ones obtained in the ICPA. In all three
symmetry directions and for all branches, the ICPA performs slightly better than the CPA for high wave-vectors. The worst agreement
with the experiment is for high wave-vectors in the $[\zeta,0,0]$ and $[\zeta,\zeta,\zeta]$ 
longitudinal branches and the $[\zeta,\zeta,\zeta]$ transverse branch. 
In these cases, the low values
of the widths in the theoretical calculations can be understood from the shape of the structure factors.  
From the examples in Fig.4 one can see that the agreement with 
experiment is
good when we have a symmetric line shape, for example, for the [.5,0,0]L mode. 
On the other hand, the
worst agreements with the experimental widths are for cases where we obtain a highly asymmetric line shape,
for example, for the [1,0,0]L mode.
Fitting Lorentzians to such asymmetric line shapes is not condusive to meaningful
values of the FWHMs. 
Because they obtain higher widths than the theories in those particular cases where the 
observations have been made with worse resolution(open circles in Fig.5), it is also
not clear how trustworthy their treatment of the resolution function is.

The discussion above clearly tells us that for Ni$_{55}$Pd$_{45}$, 
the dominant effect is mass disorder.
That alloy therefore does not provide a proper test of the ICPA. 
Nevertheless, our discussions show how a mean-crystal model can compensate
for the neglect of mass fluctuations in alloys with little force-constant disorder through the
introduction of erroneous mean force-constants, a classic case of cancellation of errors.

\section{Ni$_{50}$Pt$_{50}$; strong mass and force-constant disorder}

The mass ratio $m_{Pt}/m_{Ni}$ is 3, quite large compared to that in
the NiPd system. The force-constants of Pt are, on an average, 55$\%$ larger \cite{dbm}
than those of Ni. This makes it a potential example of strong
force-constant disorder. Tsunoda {\it et al.}\cite{expt3} investigated Ni$_{x}$Pt$_{1-x}$
by inelastic neutron scattering and compared their observations with the CPA. Here, for illustration,
we have considered $x$=0.5 only because that makes it a concentrated alloy and 
the failure of CPA was, qualitatively, very prominent at this concentration.
They compared their incoherent scattering data with that of the CPA which
predicted a split band separating out Ni and Pt contributions with a gap between
them. The experiments did not reveal a split-band, and it was very clear that
the inter-species forces play a significant role. We performed
calculations with the CPA and the ICPA. As before, we used the ICPA force-constants
in the CPA. The choice of ICPA force-constants
was more difficult than for the NiPd because of the larger 
size difference between Ni and Pt. In this alloy, the Ni-Pt
separation is also larger than the Ni-Ni separation. As a result, the Ni-Pt force-constants 
should also be less than those of Ni-Ni.
Moreover, a pair of Ni atoms would find themselves in a cage partly made of larger Pt atoms
which would therefore reduce the Ni-Ni force-constants
relative to their values in the pure material. Similarly, the bigger Pt atoms
find themselves compressed between much smaller Ni atoms, which would increase
the Pt-Pt force-constants with respect to their values in pure Pt.
Using this intuitive argument, we found that the following guesses for the force-constants
worked well:
$\phi^{xy}_{Ni-Ni}$, $\phi^{xy}_{Pt-Pt}$, $\phi^{zz}_{Ni-Ni}$ and $\phi^{zz}_{Pt-Pt}$
are kept the same as those of the pure materials \cite{dbm} and
\begin{eqnarray*}
\phi^{yy}_{Ni-Ni} = \phi^{xx}_{Ni-Ni} = 0.9 \phi^{xx}_{Ni(pure)}, \nonumber \\
\phi^{yy}_{Pt-Pt} = \phi^{xx}_{Pt-Pt} =1.1 \phi^{xx}_{Pt(pure)}, \nonumber \\
\phi^{\alpha \beta}_{Ni-Pt} = 0.8 \phi^{\alpha \beta}_{Ni-Ni}, \text{{\hspace{0.3cm}}for all $\alpha$,$\beta$.} 
\end{eqnarray*}
\begin{figure}
\includegraphics[scale=0.5]{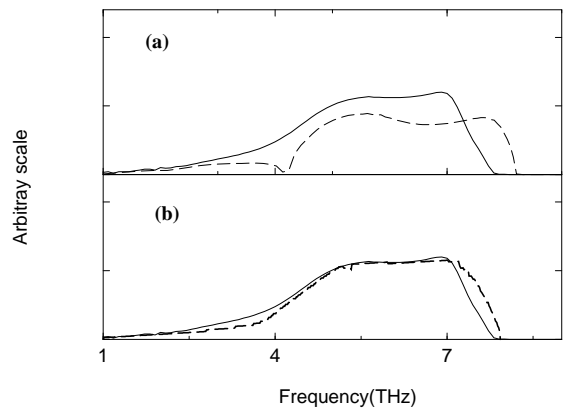}
\caption{(a)Incoherent neutron scattering structure factor vs. frequency calculated in the
ICPA(solid line) and in the CPA(dotted line) in
Ni$_{50}$Pt$_{50}$ (b) Same plot as (a) in the ICPA(solid line) and experimental
results \cite{expt3}(dotted line).}
\end{figure}
In Fig.6, we compare the ICPA results for the incoherent neutron structure factor (Eq.(54))
with those of the CPA and the experiment \cite{expt3}. In Fig. 6(a), the solid line
stands for the ICPA results while the dotted line stands for the CPA. In Fig.6(b),
the solid line is the ICPA results and the dotted line is the experimental curve.
The CPA results suggest a split-band behaviour in the middle of the band clearly
separating the Pt contribution in the low frequency region from the Ni contribution
in the high frequency region. The overall contribution from the low frequency
region is much less than that of the high frequency region in this system because the
low frequency region is dominated by the heavier atom Pt which has a much lower 
incoherent scattering length \cite{expt3} than Ni, 0.1 in comparison to 4.5 for Ni.
Including only mass fluctuations and ignoring the Ni-Pt correlated motion induced
by the environmental disorder gives rise to this spurious gap in the CPA results. This point
is further discussed below in connection with our analysis of the coherent scattering results. 
On the other hand, by incorporating the force-constant disorder, as is done in the ICPA, we
get rid of this spurious gap and obtain good agreement with the experimental results,
including the position of the right band-edge. The influence of the force-constant disorder
is demonstrated more prominently in the dispersion curves and the line shapes.
\begin{figure}
\includegraphics[scale=0.5]{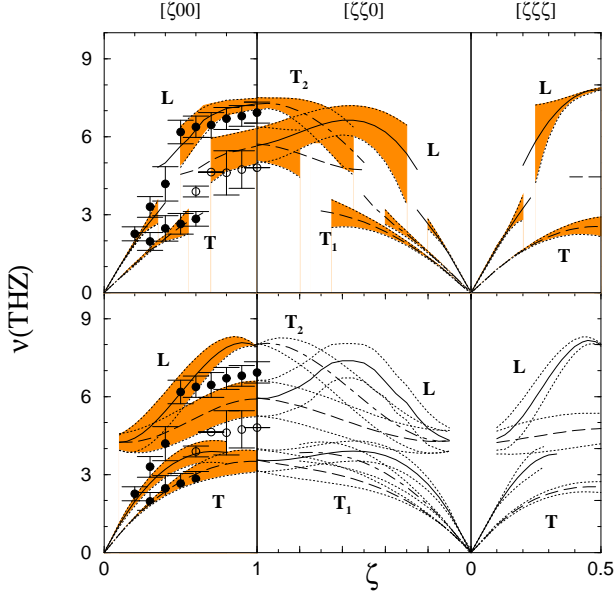}
\caption{
The solid lines are the L branch in all the three panels,
the dashed lines are the T branch in the left and the right panels.
In the central column, the long-dashed curves are the T$_{1}$ branches while
the dot-dashed curves are the T$_{2}$ branches. The shaded regions span the
FWHMs. The circles in the left panels are the experimental data
\cite{expt3}. The filled ones are those with better resolution and accuracy.
(Top panel)Dispersion curves for
Ni$_{50}$Pt$_{50}$ calculated in the ICPA.
(Bottom panel)Dispersion curves for
Ni$_{50}$Pt$_{50}$ calculated
in the CPA. Here,
the shaded regions in the left panel span the
FWHMs. In other two panels the thin dotted lines denote
the FWHMs.} 
\end{figure}
In Fig.7, we compare the dispersion curves and widths obtained in the ICPA from the
coherenet scattering structure factors, using the force
constants as given above, with those in the CPA, using the averages of the same force-constants, and
with the experimental results \cite{expt3}. The top three panels are the results obtained 
in the ICPA for the three symmetry directions. The procedure has already been discussed
in the previous section. The bottom panels are the CPA results. The circles and error bars in the
left panels are the experimental frequencies and the widths(FWHMs). The filled circles are experimental
results with more accuracy and greater resolution. The shaded regions in the top panels and
in the leftmost panel in the bottom span the calculated FWHM's. The FWHM's for the middle
and the right panels in the bottom are indicated by the thin dotted lines. For
the $[\zeta,0,0]$ and the $[\zeta, \zeta, \zeta]$ directions, the solid lines represent the
longitudinal modes and the dotted lines the transverse modes. For the $[\zeta, \zeta, 0]$
direction, the solid lines show the longitudinal mode while the long-dashed and the dot-dashed
curves stand for the T$_{1}$ and the T$_{2}$ transverse modes, respectively. The experimental
results are available only for the $[\zeta,0,0]$ directions in this system. The ICPA agrees much
better with the experiments than the CPA for both the longitudinal and the transverse branches.
The CPA frequencies are generally 
below the experimental ones at low frequencies and above the
experimental ones at high frequencies. The discrepancy gets worse as we move from the middle of the zone towards
the zone edge. This can be explained the following way:
the high wave-vector region is dominated by the lighter atoms. The use of the
average force-constants by the CPA coupled with the domination of the lighter mass pushes the frequencies 
further up, thus producing a significant deviation from the experimental ones. The severity of
this effect can be understood from the widths as well. In the CPA, the experimental points stay
well outside the disorder-induced widths centered at the peak frequencies. The discrepancy is 
substantially reduced by
the inclusion of force-constant disorder, as is seen from the ICPA results. Its inclusion
changes the dispersion curves qualitatively as well. In the CPA,
the bands extend fully across the Brilluoin zone for all symmetry directions while
in the ICPA, the Pt-dominated peaks disappear at high-$\zeta$ and the Ni-dominated
peaks wash out at low-$\zeta$ for all modes and symmetry directions, an effect
observed in the experiments. This is a clear consequence of the force-constant disorder which can be understood 
by inspecting the spectral line shapes. In Fig.8(a) we present the partial and the total spectral densities for
three different wave-vectors, one on the lower side of the zone, one in the middle and one at the boundary.
The solid lines are the total spectral functions, the dotted lines are the Ni-Ni partial spectral
functions, the long-dashed lines are the Pt-Pt contributions while the dot-dashed lines are the Ni-Pt
contributions. In Fig.8(b) we present the partial and the total coherent scattering structure factors
i.e. the spectral functions weighted by the coherent scattering lengths of the species. The coherent
scattering lengths of Ni and Pt differ by only 7$\%$(the scattering length of Ni is 1.03 while that of Pt
is 0.95), much less than do those of Ni and Pd. However, even this small difference produces significant
changes in the line shapes and in the peak frequencies. In Fig.8(b), the solid lines
give the total structure factor, the dotted lines are for weighted Ni-Ni contributions, the 
long-dashed lines give the weighted Pt-Pt contributions and the dot-dashed lines stand for weighted
Ni-Pt contributions. A close inspection of the various contributions reveals the fact that unlike in
NiPd, the Ni-Pt contribution plays the key role in determining the weight in the middle of the band
(and in obtaining the merged bands in Fig.6)
as well as adding or subtracting weights to the Ni-Ni or Pt-Pt contributions, thus elevating or
suppressing one of the peaks. For example, in Fig.8(a), in the [.5,0,0]T curves, the Ni-Pt contribution
adds weight to the total spectral function on top of the Pt-Pt peak at the low frequenies
while it subtracts weight from the Ni-Ni contribution at higher frequencies thereby causing
a weakly defined peak at high frequencies. In the [.5,0,0]L and in the [1,0,0]T curves, the Ni-Pt
contribution adds weight between the Pt-Pt and Ni-Ni peaks, thereby
removing the gap in the spectrum. The Ni-Pt contribution is totally 
due to inclusion of force-constant disorder, since diagonal disorder produces no such
contribution. Thus, the CPA produces spectral functions having two well-defined peaks
corresponding to the Ni-Ni and the Pt-Pt contributions with a gap in between resulting in
extended dispersion curves and split bands.
The effect of incorporation of the difference in scattering
lengths can be seen from these two figures as well. For example, in the [1,0,0]T curves, there are two
well-defined peaks in the total spectral functions, whereas the low-frequency peak is transformed into a shoulder
in the total structure factor. This is because Ni has the larger scattering length which enhances the
weight associated with the Ni-Pt contribution thereby cancelling more effectively the contribution from the Pt-Pt part.
Similar effects are seen in the [.5,0,0]L and [1,0,0]L curves. Moreover, this weighting sometimes
produces a weakly defined peak whose FWHM cannot be well determined, which explains the observed
washing out of the dispersion curves noted above. The effect of the small difference in scattering
lengths is amplified by the force-constant disorder through the Ni-Pt structure factor.
\begin{figure}
\includegraphics[scale=0.45]{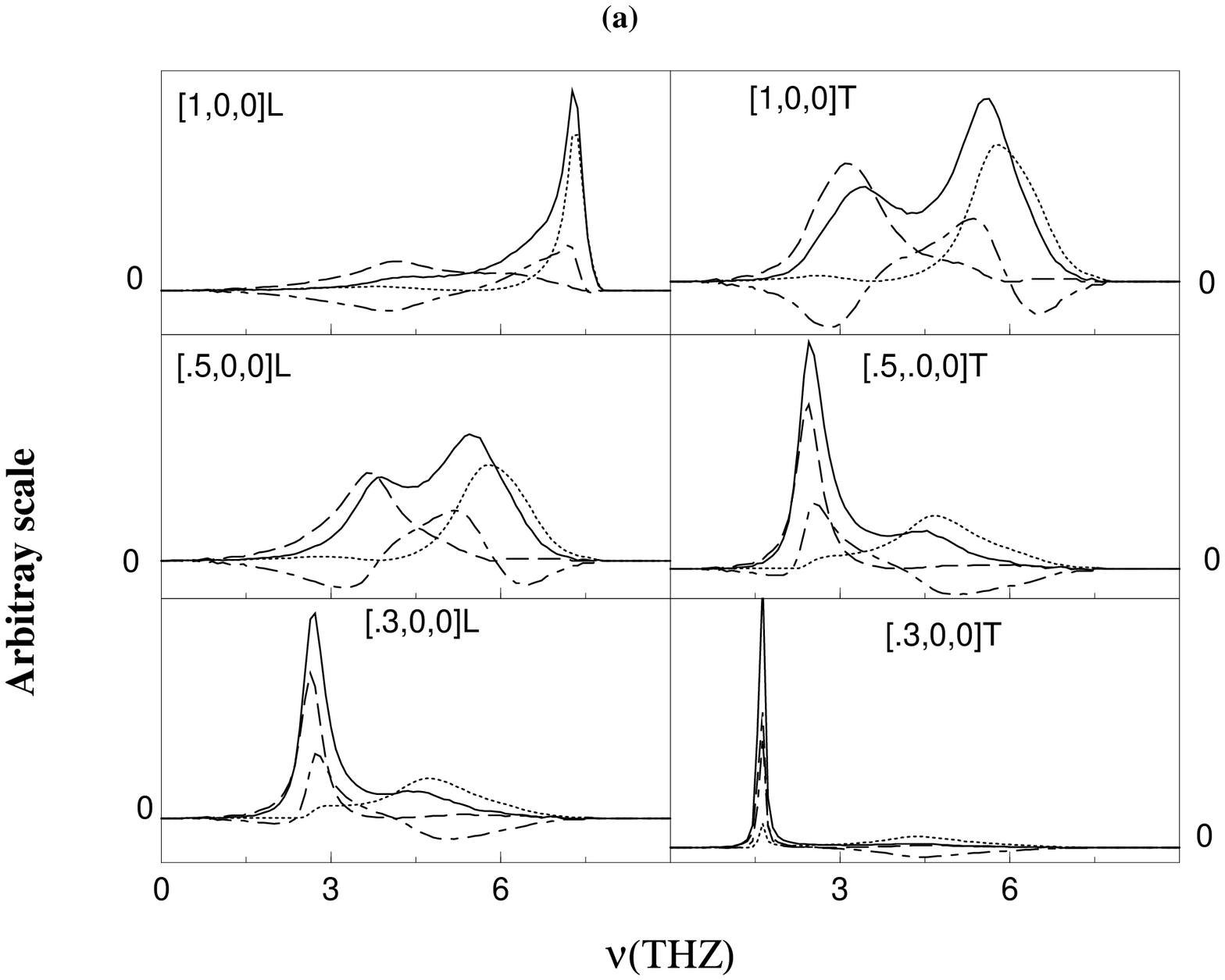}
\includegraphics[scale=0.45]{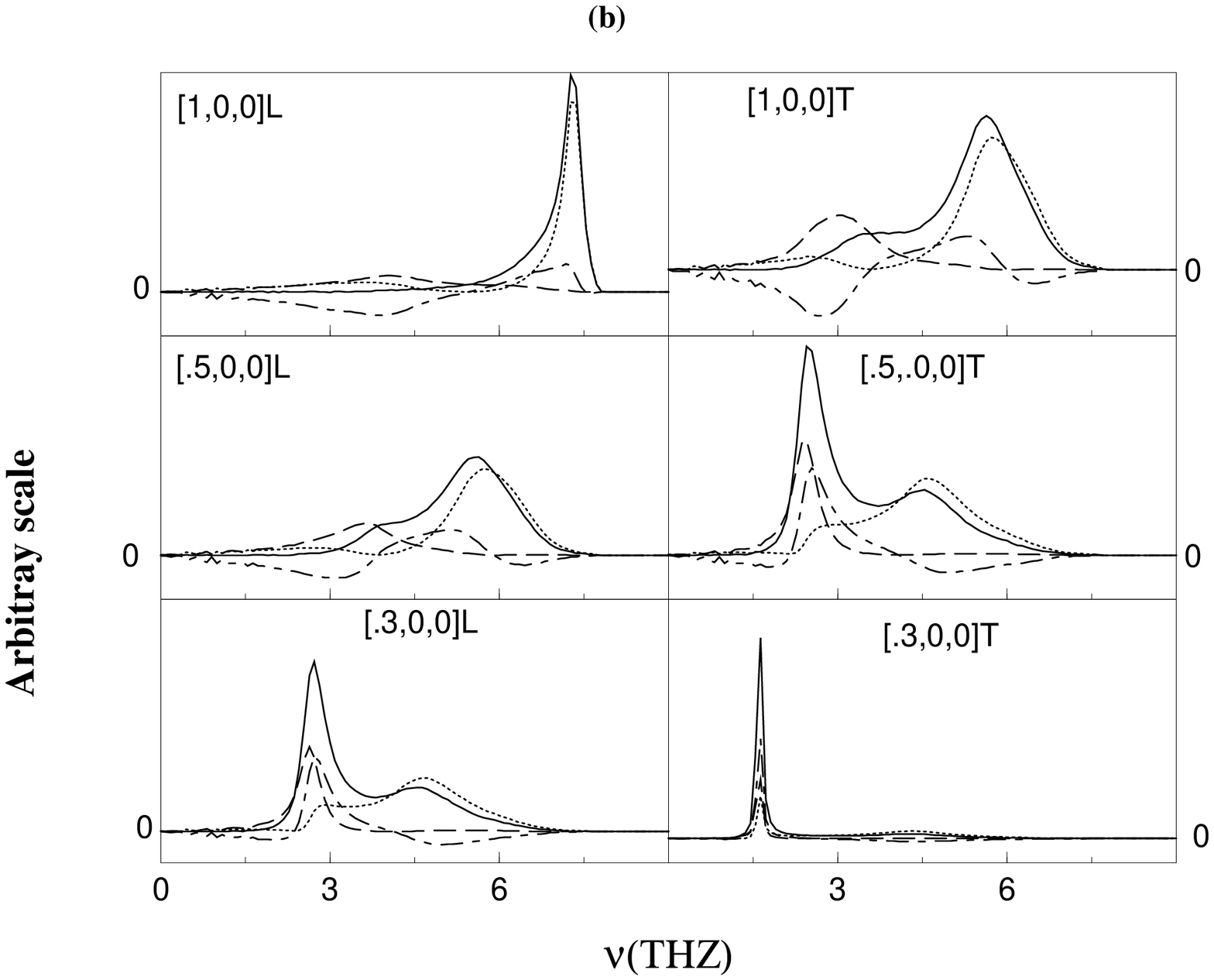}
\caption{(a)Partial and total spectral functions calculated in the ICPA 
for various $\zeta$ values in the [$\zeta$,0,0] directions in Ni$_{50}$Pt$_{50}$. (b) Partial and
total structure factors calculated in the ICPA for various $\zeta$ values in the [$\zeta$,0,0] directions 
in Ni$_{50}$Pt$_{50}$. The solid lines are the total contributions, the dotted lines
are the Ni-Ni contributions, the long-dashed lines are the Pt-Pt contributions and
the dot-dashed lines are the Ni-Pt contributions. The details are given
in the text.}
\end{figure}

In sum, the force-constant disorder plays a significant role
in Ni$_{50}$Pt$_{50}$, and a theory with mass disorder only, like the CPA, fails both qualitatively
and quantitatively in such cases. On the other hand, the ICPA successfully explains the effects
of force-constant disorder through its effect on the partial structure factors. It also demonstrates
the relative importance of the contributions of various atomic species to the coherent and
incoherent structure factors which the CPA cannot. The ICPA and the Ni$_{50}$Pt$_{50}$ 
system therefore provide a proper test case
for force-constant disorder and show that the ICPA can form a basis for understanding the lattice dynamics of other
binary alloys.

\section{Conclusions}
We have presented a straight-forward and tractable  formulation of the KLGD \cite{klgd} method for
single-site scattering of phonons in three dimensional lattices. We have demonstrated
how this multiple-scattering based formalism captures the effects of off-diagonal and environmental disorder.
The use of augmented-space
to keep track of the configurations of the system has made the formalism simple yet powerful. The resulting translational
invariance makes it numerically tractable as well. A significant contribution beyond Ref.18
is the derivation of the partial Green's functions in real-space and the derivation
of the partial spectral functions as well as their sum rules.  This enables one
to make direct comparison with neutron scattering experiments because of the
incorporation of the scattering lengths of the different species. 
We have applied the formalism to real random alloys for the first time.
In Ni$_{55}$Pd$_{45}$ we have demonstrated that mass disorder plays the prominent role, and
the CPA consequently does a rather good job whereas the mean-crystal model requires erroneous 
fitted force-constants. Our partial structure factors enable us
to understand the insensitivity of the normal modes towards the difference in the coherent scattering
lengths of the two species despite the significant difference of 43$\%$ in this system. The Ni$_{50}$Pt$_{50}$
results demonstrate the prominence of force-constant disorder even in a case where the mass ratio is 3.
The ICPA agrees well with both the coherent and the incoherent scattering 
experiments, whereas the CPA fails, both quantitatively and qualitatively. We are able to establish
the role of the force-constant differences between the species in great detail with the help of
the partial spectral functions and the partial structure factors. We have consequently clearly demonstrated that for
systems like NiPt, where the force-constants are strongly species dependent, the
determination of their values is crucial. However, we had no prior information about the 
species dependence of the force-constants.
Intuitive arguments led to a set of force-constants
which turned out to be quite good. A better understanding of the role of disorder
in the lattice dynamics of random alloys could be achieved with
prior information about the force-constants. These could  be obtained, e.g., from first principles calculations on 
a set of ordered alloys.

\acknowledgments
We thank Professors David Vanderbilt, Gabriel Kotliar and Karin Rabe 
for useful discussions and for the use of their
computer systems. We thank W. Kamitakahara and R. Nicklow for useful communications
regarding their experimental results.
\newpage
\appendix
\section{Matrix elements}

For the calculations in the nearest-neighbor approximation, one needs to evaluate the $\{3\left(Z+1 \right)\}^{2}$
matrix elements of the operators ${\bf K}^{\prime}$, ${\bf K}^{\prime \dagger}$, 
$\widetilde{\bf V}\left(\vec{q} \right)$ and ${\bf G}_{vca}^{-1}$ and use them as inputs. These evaluations are done
in augmented space using Eq.(14). The symmetry of the lattice
structure is used to reduce the number of matrix elements evaluated. 
Here, we give results
only for an fcc lattice. All the matrices are, therefore, of dimension $39 \times 39$.

In an fcc system, each atom has 12 nearest neighbors with coordinates $\left(\pm \frac{1}{2}, \pm \frac{1}{2},0 \right)$, $\left(\pm \frac{1}{2},0,\pm \frac{1}{2} \right)$, and 
$\left(0,\pm \frac{1}{2}, \pm \frac{1}{2} \right)$ with respect to the coordinates of the
reference atom at $\left(0,0,0 \right)$. 
The force-constants, between the atom $0$ and its neighbors satisfy the following 
cubic symmetry relation.
\begin{equation}
\phi^{\alpha \beta}_{0j}= \phi^{\beta \alpha}_{0j}= \phi^{\alpha \beta}_{j0}=\phi^{\alpha \beta}_{0k}, 
\end{equation}
where
$\vec{R}_{0j}=-\vec{R}_{0k}$ and
$k$ and $j$ are two neighbors on opposite sides of site $0$. 
For example, the force-constant matrix
between the atoms $\left(0,0,0 \right)$ and $\left(\frac{1}{2},\frac{1}{2},0 \right)$ is of the form,
\begin{equation}
\phi_{(000,\frac{1}{2}\frac{1}{2}0)}= \left( \begin{array} {ccc}
	a & b & 0 \\ b & a & 0 \\ 0 & 0 & g
	\end{array} \right) .
\end{equation}
The force-constant matrices between the atom $0$ and its other neighbors can easily be calculated
from (A2) via the cubic symmetry operations.
The results are:
\begin{widetext}
\begin{eqnarray}
\left(G_{vca}^{-1} \right)^{\alpha\beta}_{ij} &=&
	\bar{m}\omega^{2}-8D_{1}^{xx}-4D_{1}^{\prime xx}\text{\hspace {2.0cm}} \text{if $i=j$, $\alpha=\beta$}, \nonumber\\
	& = & 0 \text{\hspace {5.1cm}} \text{if $i=j$, $\alpha \neq \beta$}, \nonumber \\
	& = & D_{1}^{\prime xx}  \text {\hspace {4.5cm}}\text{ if $i=0$, $j=n=1-12$, $\alpha= \beta$, $R^{\alpha}_{j}=0$} , \nonumber\\
	& = & D_{1}^{xx} \text{\hspace {4.7cm}} \text{if $i=0$, $j=n=1-12$, $\alpha= \beta$, $R^{\alpha}_{j}\neq 0$ }, \nonumber \\
	& = & 2 R^{\alpha}_{j}\times 2R^{\beta}_{j}\times D_{1}^{xy} 	\text{\hspace {2.6cm}} \text{if $i=0$, $j=n=1-12$, $\alpha \neq \beta$}, \nonumber \\
& = & 0 \text{\hspace{5.2cm}} \text{otherwise},
\end{eqnarray}
and
\begin{eqnarray*} 
\left(G_{vca}^{-1} \right)^{\alpha\beta}_{ij} = \left(G_{vca}^{-1} \right)^{\alpha\beta}_{ji}= \left(G_{vca}^{-1} \right)^{\beta\alpha}_{ij}, \text{\hspace{0.5cm}}\text{for all $i$,$j$,$\alpha$, and $\beta$}.  
\end{eqnarray*}
Also, we find
\begin{eqnarray}
\left(K^{\prime} \right)^{\alpha\beta}_{ij} &=&
	m^{\prime}\omega^{2}-8D_{2}^{xx}-4D_{2}^{\prime xx}\text{\hspace {2.0cm}} \text{if $i=j=0$, $\alpha=\beta$}, \nonumber\\
	& = & 0 \text{\hspace {5.2cm}} \text{if $i=j=0$, $\alpha \neq \beta$}, \nonumber \\
	& = & D_{2}^{\prime xx}  \text {\hspace {4.5cm}}\text{ if $i=0$, $j=n=1-12$, $\alpha= \beta$, $R^{\alpha}_{j}=0$} , \nonumber\\
	& = & D_{2}^{xx} \text{\hspace {4.7cm}} \text{if $i=0$, $j=n=1-12$, $\alpha= \beta$, $R^{\alpha}_{j}\neq 0$ }, \nonumber \\
	& = & 2 R^{\alpha}_{j}\times 2R^{\beta}_{j}\times D_{2}^{xy}\text{\hspace {2.6cm}} \text{if $i=0$, $j=n=1-12$, $\alpha \neq \beta$}, \nonumber \\
& = & -\left(K^{\prime} \right)^{\alpha\beta}_{0j} \text{\hspace {3.9cm}} \text{if $i \neq 0$, $j=i=n=1-12$}, \nonumber \\
& = & 0 \text{\hspace{5.2cm}} \text{otherwise},
\end{eqnarray}
and
\begin{eqnarray*} 
\left(K^{\prime} \right)^{\alpha\beta}_{ij} = \left(K^{\prime} \right)^{\alpha\beta}_{ji}= \left(K^{\prime} \right)^{\beta\alpha}_{ij}, \text{\hspace{0.5cm}}\text{for all $i$,$j$,$\alpha$, and $\beta$}.  
\end{eqnarray*}
Similarly, we obtain
\begin{eqnarray}
\widetilde{V}\left(\vec{q} \right) ^{\alpha \beta}_{ij} & = &
L^{\alpha} \text{\hspace{5.6cm}} \text{if $i=j=0$, $\alpha=\beta$}, \nonumber \\
&=& 4D_{3}^{xy}Sin(q_{\alpha})Sin(q_{\beta}) \text{\hspace{2.8cm}} \text{if $i=j=0$, $\alpha \neq \beta$}, \nonumber \\
	& = & D_{5}  \text {\hspace {5.4cm}}\text{ if $i=0$, $j=n=1-12$, $\alpha= \beta$, $R^{\alpha}_{j}=0$} , \nonumber\\
	& = & D_{6} \text{\hspace {5.5cm}} \text{if $i=0$, $j=n=1-12$, $\alpha= \beta$, $R^{\alpha}_{j}\neq 0$ }, \nonumber \\
	& = & 2 R^{\alpha}_{j}\times 2R^{\beta}_{j}\times D_{7}	\text{\hspace {3.4cm}} \text{if $i=0$, $j=n=1-12$, $\alpha \neq \beta$}, \nonumber \\
	& = & D_{4}^{\prime xx}-D_{1}^{\prime xx}  \text {\hspace {4.0cm}}\text{ if $i \neq 0$, $j=i=n=1-12$, $\alpha= \beta$, $R^{\alpha}_{i}=0$} , \nonumber\\
	& = & D_{4}^{xx}-D_{1}^{xx} \text{\hspace {4.3cm}} \text{if $i \neq 0$, $j=i=n=1-12$, $\alpha= \beta$, $R^{\alpha}_{i}\neq 0$ }, \nonumber \\
	& = & 2 R^{\alpha}_{j}\times 2R^{\beta}_{j}\times \left(D_{4}^{xy}-D_{1}^{xy} \right) 	\text{\hspace {1.8cm}} \text{if $i \neq 0$, $j=i=n=1-12$, $\alpha \neq \beta$}, \nonumber \\
& = & - D_{3}^{\prime xx}e^{i\vec{q}\cdot \vec{R}_{i}} \text{\hspace{4.2cm}} \text{if $j \neq i=n=1-12$, $\vec{R}_{j}=-\vec{R}_{i}$, $\alpha=\beta$, $R_{j}^{\alpha}=0$}, \nonumber \\
& = & - D_{3}^{xx}e^{i\vec{q}\cdot \vec{R}_{i}} \text{\hspace{4.3cm}} \text{if $j \neq i=n=1-12$, $\vec{R}_{j}=-\vec{R}_{i}$, $\alpha=\beta$, $R_{j}^{\alpha} \neq 0$}, \nonumber \\
& = & - D_{3}^{xy}e^{i\vec{q}\cdot \vec{R}_{i}}\times 2R_{i}^{\alpha}\times 2R_{i}^{\beta} \text{\hspace{2.2cm}} \text{if $j \neq i=n=1-12$, $\vec{R}_{j}=-\vec{R}_{i}$, $\alpha \neq \beta$}, \nonumber \\
& = & 0 \text{\hspace{5.8cm}}\text{otherwise},
\end{eqnarray}
and
\begin{eqnarray*} 
\widetilde{V}\left(\vec{q} \right) ^{\alpha \beta}_{ij} &=& \left(\widetilde{V}\left(\vec{q} \right)^{\star}\right)^{\alpha \beta}_{ji}= \widetilde{V}\left(\vec{q} \right)^{\beta \alpha}_{ij}, \text{\hspace{0.5cm}}\text{for all $i$,$j$,$\alpha$, and $\beta$}. 
\end{eqnarray*}
In these evaluations, we have used the notation $R_{0j}=R_{j}$, $n=1-12$ to represent the 12 nearest neighbors, and the notations
\begin{eqnarray}
\bar{m} & = & c_{A}m^{A}+ c_{B}m^{B} , \nonumber \\
m^{\prime} & = & \sqrt{c_{A}c_{B}}\left(m^{A}-m^{B} \right) , \nonumber \\
\tilde{m} & = & c_{B}m^{A}+ c_{A}m^{B} , \nonumber \\
D_{1}^{xx} & = & c_{A}^{2}a_{AA} + c_{B}^{2}a_{BB} +2c_{A}c_{B}a_{AB} , \nonumber \\
D_{1}^{\prime xx} & = & c_{A}^{2}g_{AA}+c_{B}^{2}g_{BB} +2c_{A}c_{B}g_{AB} , \nonumber \\
D_{1}^{xy} & = & c_{A}^{2}b_{AA}+c_{B}^{2}b_{BB} +2c_{A}c_{B}b_{AB} , \nonumber \\
D_{2}^{xx} & = & \sqrt{c_{A}c_{B}}\{c_{A}a_{AA} -c_{B}a_{BB} +(c_{B}-c_{A})a_{AB} \} , \nonumber \\
D_{2}^{\prime xx} & = & \sqrt{c_{A}c_{B}}\{c_{A}g_{AA} -c_{B}g_{BB} +(c_{B}-c_{A})g_{AB} \} , \nonumber \\
D_{2}^{xy} & = & \sqrt{c_{A}c_{B}}\{c_{A}b_{AA} -c_{B}b_{BB} +(c_{B}-c_{A})b_{AB} \} , \nonumber \\
D_{3}^{xx} & = & c_{A}c_{B} \left(a_{AA} + a_{BB} -2a_{AB} \right) , \nonumber \\
D_{3}^{\prime xx} & = & c_{A}c_{B} \left(g_{AA} + g_{BB} -2g_{AB} \right) , \nonumber \\
D_{3}^{xy} & = & c_{A}c_{B} \left(b_{AA} + b_{BB} -2b_{AB} \right) , \nonumber \\
D_{4}^{xx} & = & c_{A}c_{B} \left(a_{AA}+a_{BB} \right) + \left(c_{A}^{2}+c_{B}^{2} \right)a_{AB} , \nonumber \\
D_{4}^{\prime xx} & = & c_{A}c_{B} \left(g_{AA}+g_{BB} \right) + \left(c_{A}^{2}+c_{B}^{2} \right)g_{AB} , \nonumber \\
D_{4}^{xy} & = & c_{A}c_{B} \left(b_{AA}+b_{BB} \right) + \left(c_{A}^{2}+c_{B}^{2} \right)b_{AB} , \nonumber \\
D_{5} & = & \left(D_{1}^{xx}-D_{4}^{xx} \right) + D_{3}^{xx} e^{i \vec{q}\cdot \vec{R}_{j}} , \nonumber \\
D_{6} & = & \left(D_{1}^{xy}-D_{4}^{xy} \right) + D_{3}^{xy} e^{i \vec{q}\cdot \vec{R}_{j}} , \nonumber \\
D_{7} & = & \left(D_{1}^{\prime xx}-D_{4}^{\prime xx} \right) + D_{3}^{\prime xx} e^{i \vec{q}\cdot \vec{R}_{j}} , \nonumber \\
L^{\alpha} & = & \left(\bar{m}-\tilde{m} \right)\omega^{2} - 8\left(D_{1}^{xx} - D_{4}^{xx} \right) - 4\left(D_{1}^{\prime xx} - D_{4}^{\prime xx} \right) - 4D_{3}^{xx} \{Cos\enskip q_{\alpha}\left(Cos\enskip q_{\gamma}+Cos\enskip q_{\delta} \right) \nonumber \\
& & \text{\hspace{4.0cm}}- 4D_{3}^{\prime xx} Cos\enskip q_{\gamma} Cos\enskip q_{\delta}; \text{\hspace{2.0cm}}\gamma,\delta \neq \alpha .
\end{eqnarray}
\end{widetext}
The symmetries of the force-constant matrices are reflected in the operators as
well. The effect of {\it itineration} is captured in $\widetilde{V}\left(\vec{q}\right)^{\alpha \beta}_{ij}$
through the quantities $D_{3}^{\alpha \beta}$. 
When there is no force-constant disorder the $D_{3}$ terms vanish, 
and $\widetilde{\bf V}$ becomes independent of $\vec{q}$. A $\vec{q}$-independent self-energy 
results, and we arrive at the CPA equations.
To illustrate how to obtain the various matrix elements of the operators, we present the
calculation of $K^{\prime (0)}_{01}$ where $R_{1}=\left(\frac{1}{2}, \frac{1}{2}, 0 \right)$.
\begin{eqnarray*}
K^{\prime (0)}_{01} & = & \langle 0f \vert \widehat{\bf K} \vert 1f_{0} \rangle , \nonumber \\
& = & \langle 0f \vert \left(\widehat{\bf m}\omega^{2} - \widehat{\bf \Phi} \right) \vert 1f_{0} \rangle.
\end{eqnarray*}
Using Eqs.(6)and (11),
\begin{eqnarray*}
\langle 0f \vert \widehat{\bf m} \vert 1f_{0} \rangle & = & 0 ,
\end{eqnarray*}
and using Eqs.(7) and (11),
\begin{widetext}
\begin{eqnarray*}
\langle 0f \vert \widehat{\bf \Phi} \vert 1f_{0} \rangle & = & \langle f \vert \widehat{\bf \Phi}_{01} \vert f_{0} \rangle , \nonumber \\
& = & \langle \left(\sqrt{c_{A}}\langle A_{0} \vert + \sqrt{c_{B}} \langle B_{0} \vert \right), \left(\sqrt{c_{A}} \langle A_{1} \vert + \sqrt{c_{B}} \langle B_{1} \vert \right) \vert \{\phi_{01}^{AA} \widehat{\bf \eta}_{0}^{A} \widehat{\bf \eta}_{1}^{A} + \phi_{01}^{BB} \widehat{\bf \eta}_{0}^{B} \widehat{\bf \eta}_{1}^{B} + \phi_{01}^{AB} \widehat{\bf \eta}_{0}^{A} \widehat{\bf \eta}_{1}^{B} + \phi_{01}^{BA} \widehat{\bf \eta}_{0}^{B} \widehat{\bf \eta}_{1}^{A} \}\vert \nonumber \\
& &  \text{\hspace{5.0cm}}\left(\sqrt{c_{B}}\vert A_{0} \rangle - \sqrt{c_{A}} \vert B_{0} \rangle \right) , \left(\sqrt{c_{A}} \vert A_{1} \rangle + \sqrt{c_{B}} \vert B_{1} \rangle \right)\rangle.
\end{eqnarray*}
\end{widetext}
If we use the cartesian coordinates explicitly, then the xx component is, for example, 
\begin{eqnarray*}
\langle f \vert \Phi^{xx}_{01} \vert f_{0} \rangle & = & \sqrt{c_{A}c_{B}} \{ c_{A}a_{AA} -c_{B}a_{BB} + (c_{B}-c_{A})a_{AB} \} \nonumber \\
& = & D_{2}^{xx} .
\end{eqnarray*}
The other components can be calculated similarly.

\end{document}